\documentclass[11pt]{article}
\usepackage[table]{xcolor}
\usepackage{graphicx} 
\usepackage{style}
\usepackage[framemethod=tikz]{mdframed}
\title{Unbiased Insights: Optimal Streaming Algorithms for $\ell_p$ Sampling, the Forget Model, and Beyond}
\author{Honghao Lin\footnote{(honghaol@andrew.cmu.edu) Computer Science Department, Carnegie Mellon University. Supported in part by a Simons Investigator Award and a CMU Paul and James Wang Sercomm Presidential Graduate Fellowship.
\smallskip} 
\\ {\small Carnegie Mellon University} 
\and
Hoai-An Nguyen\footnote{(hnnguyen@andrew.cmu.edu) Computer Science Department, Carnegie Mellon University.
Supported in part by an NSF GRFP fellowship grant number DGE2140739 and NSF CAREER Award CCF-2330255. \smallskip} \\ {\small Carnegie Mellon University} 
\and
William Swartworth\footnote{(wswartworth@gmail.com) Computer Science Department, Carnegie Mellon University.
Supported in part by Office of Naval Research award number N000142112647 and a Simons Investigator Award. \smallskip} \\ {\small Carnegie Mellon University} 
\and
David P. Woodruff\footnote{(dwoodruf@cs.cmu.edu) Computer Science Department, Carnegie Mellon University. Supported in part by Office of Naval Research award number N000142112647 and a Simons Investigator Award.\smallskip} \\ {\small Carnegie Mellon University} }
\date{}

\begin{document}

\maketitle

\begin{abstract}
We study $\ell_p$ sampling and frequency moment estimation in a single-pass insertion-only data stream. For $p \in (0,2)$, we present a nearly space-optimal approximate $\ell_p$ sampler that uses $\widetilde{O}(\log n \log(1/\delta))$ bits of space and for $p = 2$, we present a sampler with space complexity $\widetilde{O}(\log^2 n \log(1/\delta))$. This space complexity is optimal for $p \in (0, 2)$ and improves upon prior work by a $\log n$ factor. We further extend our construction to a continuous $\ell_p$ sampler, which outputs a valid sample index at every point during the stream.

Leveraging these samplers, we design nearly unbiased estimators for $F_p$ in data streams that include forget operations, which reset individual element frequencies and introduce significant non-linear challenges. As a result, we obtain near-optimal algorithms for estimating $F_p$ for all $p$ in this model, originally proposed by Pavan, Chakraborty, Vinodchandran, and Meel [PODS'24], resolving all three open problems they posed.

Furthermore, we generalize this model to what we call the suffix-prefix deletion model, and extend our techniques to estimate entropy as a corollary of our moment estimation algorithms. Finally, we show how to handle arbitrary coordinate-wise functions during the stream, for any $g \in \mathcal{G}$, where $\mathcal{G}$ includes all (linear or non-linear) contraction functions.
\end{abstract}
\newpage
\section{Introduction}
The \emph{streaming} model of computation, motivated by the need to process massive datasets in applications such as network monitoring and real-time analytics, has been widely studied. In this model, an underlying frequency vector $\ff \in \Z^n$, initially set to the all-zero vector, is updated over time by a stream of operations. These operations typically consist of coordinate-wise increments (and sometimes decrements). Formally, the stream consists of updates of the form $(i,w_t)$, meaning $f_i\gets f_i + w_t$. When  $w_t$  can be both positive and negative, the model is referred to as a turnstile stream. In contrast, when  $w_t$  can only be positive, the model is referred to as an insertion-only stream.

A central problem in this setting is the estimation of \emph{frequency moments}, defined as $F_p = \sum_{i \in [n]} f_i^p$ for a parameter $p \geq 0$.  The (common) goal is to compute a multiplicative $(1 \pm \eps)$-approximation to $F_p$ using a constant number of passes over the stream and sublinear space. This problem has been studied in various streaming models, including insertion-only, turnstile, and random-order streams since the work of Alon, Matias, and Szegedy~\cite{AMS1999space}. Frequency moments have a wide range of applications. For instance, $F_0$---the number of distinct elements---arises in query optimization and database systems \cite{HNSS1995sampling}; $F_1$
corresponds to approximate counting \cite{NY2022optimal}; higher moments $F_p$ for $p \geq 2$
capture data skew and have been used in algorithms for data partitioning \cite{DNSS1992practical} and error estimation \cite{IP1995balancing}.

A closely related and extensively studied problem is \emph{$\ell_p$ sampling}. Here, given the frequency vector $\ff \in \Z^n$, the goal is to return an index $i \in \{1,2,\ldots, n\}$ with probability $|f_i|^p / \| \ff \|_p^p$. In particular, an approximate  $\ell_p$ sampler is one that returns an index $i$ with probability $(|f_i^p| / \| \ff \|_p^p)  \cdot (1 \pm \eps) + \poly(1/n)$ for $\eps \in (0,1)$, while a perfect $\ell_p$ corresponds to the case $\eps = 0$. $\ell_p$ samplers have proved an important tool in the streaming model and have been used in algorithms to estimate the $F_p$ moment, finding heavy hitters, finding duplicates in streams, and cascaded norm estimation \cite{MW2010pass, AKO2010streaming, JST2011tight, BOZ2012optimal}. See \Cref{intro:RL} for a more extensive overview of previous works for both $F_p$ moment estimation and $\ell_p$ sampling. 

\paragraph{$\ell_p$ Sampling.} The work of Jayaram and Woodruff~\cite{JW2021perfect} presents perfect $\ell_p$ samplers for turnstile streams that use $\widetilde{O}(\log^2 n \log(1/\delta))$ bits of space for $p \in (0,2)$ and $O(\log^3 n \log(1/\delta))$ bits for $p = 2$. On the other hand, Kapralov, Nelson, Pachocki, Wang, and Woodruff~\cite{KNPWWY2017optimal} establishes a lower bound of $\Omega( \log^2 n\log(1/\delta))$ for $\ell_p$ sampling in turnstile streams, implying that the sampler of~\cite{JW2021perfect} is tight up to a $\operatorname{poly}(\log \log n)$ factor when $p \in (0, 2)$.
However, the situation for insertion-only streams remains less clear. A trivial lower bound of $\Omega(\log n)$ is known since that many bits is required to output an index, but no upper bound better than that of~\cite{JW2021perfect} has been established even in this restricted model. This gap naturally leads to the following question:
\begin{quote}
    \em {Question 1: Is it possible to design a one-pass $\ell_p$ sampling algorithm that uses $\widetilde{O}(\log n)$ space in a insertion-only stream?}
\end{quote}

\paragraph{Forget Model.}
As mentioned, a typical application of $\ell_p$ sampling is the estimation of the frequency moment $F_p$, a problem extensively studied in both insertion-only and turnstile streaming models. To better reflect practical needs and to explore the boundaries of sketching-based estimation, in this work we consider a more general class of nonlinear update operations—specifically, {\em forget} requests.

This model, introduced by Pavan, Chakraborty, Vinodchandran, and Meel~\cite{PCVM2024feasibility}, permits updates that reset individual coordinates of the frequency vector to zero. Forget updates are motivated by several practical scenarios, including privacy-preserving deletion (e.g., to comply with the European Union's General Data Protection Regulation, which grants individuals the ``Right to be Forgotten"~\cite{PCVM2024feasibility}), storage optimization via removal of stale data, and the ability to discard faulty or irrelevant information without restarting the stream. Additionally, forget operations enable post hoc analysis on subsets of the data—for example, by zeroing out coordinates that fall outside a region of interest after the stream concludes.
 
Unlike decrements, forget operations cannot be simulated in the turnstile model without access to the current value of the coordinate being reset. This added nonlinearity introduces substantial challenges: standard linear sketching techniques become too biased and are no longer suitable for accurate $F_p$ estimation. As a result, designing unbiased or low-bias estimators in the presence of such operations requires fundamentally new techniques beyond classical streaming tools.

It is known that no sublinear-space algorithm can provide a $(1 \pm \varepsilon)$-approximation to the $F_p$ moments under unrestricted forgets. In particular, \cite{PCVM2024feasibility} established an $\Omega(n)$ space lower bound for this setting. To overcome this barrier, they proposed the $\alpha$-RFDS model, in which the algorithm is guaranteed that the final value of the $F_p$ moment is at least a $(1 - \alpha)$ fraction of what it would have been had no forgets occurred. This permits, for example, the moment to decrease by an arbitrarily large constant factor, as long as the overall loss is bounded.

Under this promise, the work of \cite{PCVM2024feasibility} designed $(1\pm \eps)$-approximation algorithms for estimating $F_0$ using $O\left(\frac{1}{\eps^2(1-\alpha) }\log n\right)$ bits of space and for $F_1$ using $O\left(\frac{1}{\eps^2(1-\alpha) }\log^2 n\right)$ bits of space. They also posed several open questions regarding their optimality and the feasibility of extending such results to other $F_p$ moments, such as $F_2$:

\begin{quote}
    {\em Question 2: What is the tight space complexity for  $F_0$ and $F_1$ estimation in the $\alpha$-RFDS model? 
    
    Question 3: Can we design space-efficient algorithms for estimating other $F_p$ moments in this model?}
\end{quote}

\paragraph{More Extensions.} 
Beyond answering the the above three questions, we consider several extensions. We introduce a \emph{prefix-suffix deletion} model, which strictly generalizes forget operations by restricting deletions based on time. A prefix deletion request of the form $\texttt{prefixdelete}(i, x)$ removes all occurrences of element $x$ that appeared at or before time $i$, while a suffix deletion request $\texttt{suffixdelete}(i, x)$ removes all occurrences of $x$ from time $i$ up to the current time $T$ when the request is received. This model allows fine-grained control over the retained data and captures a broader range of post hoc filtering operations.
For example, at the end of the stream, one can use prefix and suffix deletions to zoom in on a subset of elements in a particular time interval. 
The model is also closely related to the sliding window paradigm in streaming algorithms \cite{DGIM2002maintaining, BGLWZ2018nearly}. As an illustration, suppose we want to compute a statistic over a sliding window of size $w$; we can run overlapping sub-streams of length $2w$ every $w$ steps, and apply appropriate prefix deletions to recover the exact window.
We believe that prefix and suffix deletions---as well as forget requests---model a wide range of realistic scenarios where data retention is limited by time, privacy, or context-specific relevance.

Moreover, since supporting only forget operations may not adequately capture practical scenarios, we initiate a broader study of non-linear operations in insertion-only streams, significantly generalizing the forget model.
 Specifically, we consider a family of functions $\mathcal{G}$ such that stream updates may take the form $(i, g)$ for $g \in \mathcal{G}$, which applies the transformation $f_i \gets g(f_i)$. The function family $\mathcal{G}$ includes all contraction functions, encompassing both linear and nonlinear transformations—such as $\sqrt{x}$ and $x/c$ for a constant $c$—that reduce the magnitude of their input. See \Cref{subsec:GO} for the formal definition of $\mathcal{G}$.
Many functions in this family are nonlinear (as in the case of forgets), and thus introduce new challenges in designing accurate estimators. As in the $\alpha$-RFDS model, we assume that these operations do not reduce the final value of the $F_p$ moment by more than a $(1 - \alpha)$ multiplicative factor compared to the version of the stream with only insertions. 

\subsection{Our Contributions}
In this paper, we resolve all of the above open questions.

\paragraph{$\ell_p$ Sampling for $p \in (0,2]$.} We present an $\ell_p$ sampler for insertion-only streams that succeeds with  probability at least $1-\delta$ using $\widetilde{O}(\log n \log(1/\delta))$ bits of space for $p \in (0,2)$ and $\widetilde{O}(\log^2 n \log(1/\delta))$ bits for $p = 2$. 

\begin{theorem}[$\ell_p$ sampler]
    \label{thm: sampler_lp}
    For any constant $p \in (0, 2]$, there is a one-pass streaming algorithm that runs in space $\widetilde{O}(\log^{c(p)} n \log(1/\delta))$, and outputs an index $i$ such that with probability at least $1 - \delta$ we have for every $j \in [n]$,
    \[\operatorname{Pr}[i = j]=\frac{\left|f_j\right|^p}{\|\ff\|_p^p} \pm \frac{1}{\poly{(n)}} \;. \]
    Here $c(p) = 1$ when $0 < p < 2$ and $c(p) = 2$ when $p = 2$.
\end{theorem}

We note that \Cref{thm: sampler_lp} improves upon the results of~\cite{JW2021perfect} by a factor of $\log n$ for insertion-only streams. For $p \in (0, 2)$, our $\ell_p$ sampler achieves {\em optimal} space complexity up to a $\poly(\log \log n)$ factor. Unlike the {\em perfect} sampler from~\cite{JW2021perfect}, ours does not explicitly output \texttt{FAIL} upon failure. However, when the required failure probability is $\delta = \poly(1/n)$ for both samplers, our sampler retains all the desirable properties of the perfect sampler in~\cite{JW2021perfect} while still having an $O(\log n)$ space savings.

We next extend our $\ell_p$ sampler to the continuous case where we are required to have a sampled index at each time step during the stream. 
\begin{theorem}[Continuous $\ell_p$ sampler]
    Consider a stream of length $m$ (and therefore with $m$ time steps). Let $F_{p, [0,t]}$ be the $F_p$ moment of $\ff$ after the first $t$ time steps, and let $f_{i, [0,t]}$ be the frequency of coordinate $i$ after the first $t$ time steps. 
    There is an algorithm $\mathcal{A}$ that produces a series of $m$ outputs, $z_1, \ldots, z_m$ such that each $z_j \in \{1, \ldots, n\}$, one at each time step with the following properties:
    \begin{itemize}
        \item $\Pr(z_t = i) = \frac{f^p_{i, [0,t]}}{F_{p,[0,t]}} \pm \frac{1}{\poly(n)}$.
        \item Only $\widetilde{O}(\log^{c(p)} n \cdot \log(1/\delta))$ of the $z_j$'s are unique. 
        \item $\mathcal{A}$ uses $\widetilde{O}(\log^{c(p)} n \cdot  \log(1/\delta))$ bits of space.
        \item $\mathcal{A}$ succeeds with probability at least $1-\delta$. 
    \end{itemize}
    Here $c(p) = 1$ from $0 < p < 2$ and $c(p) = 2$ for $p = 2$.
\end{theorem}

\paragraph{$F_p$ Estimation with Forgets.} 
We give nearly-optimal algorithms for estimating $F_p$ moment in the $\alpha$-RFDS model for all range of $p$. 

\begin{theorem}[Main result for $F_p$ estimation with forgets]
    Given $0 < p < \infty$. There is a one-pass streaming algorithm that uses $m$ bits of space and with high constant probability outputs a $(1 \pm \eps)$-approximation to the value of $F_p$
in the $\alpha$-RFDS model, where
\[m = 
\begin{cases}
    \widetilde{O}\left(\frac{1}{\eps^2(1 - \alpha)} \cdot \log n\right), &p \in (0, 2), \\
    \widetilde{O}\left(\frac{1}{\eps^2(1 - \alpha)} \cdot \log^2 n\right), &p = 2, \\
    \widetilde{O}\left(\frac{1}{\eps^2(1 - \alpha)^{2/p}} \cdot n^{1 - 2/p}\cdot \log^2 n\right), &p \in (2, \infty).
 \end{cases}    
\]
\end{theorem} \label{thm:fpestmain}

We also provide matching lower bounds.
\begin{theorem}[Lower bounds for $F_p$ estimation with forgets]
    Any streaming algorithm that gives a $(1\pm \eps)$-approximation to the $F_p$ moment in the $\alpha$-RFDS model requires $\Omega\left(\frac{1}{\eps^2(1 - \alpha)} \cdot \log n\right)$ bits of space for $p \in [0, 2]$ and $\Omega\left(\frac{1}{\eps^2 (1-\alpha)^{2/p}} \cdot n^{1 - 2/p}\right)$ bits of space for $p > 2$.
\end{theorem}

We summarize our results in \Cref{tab:1}. In addition, we give a surprisingly simple algorithm that achieves the optimal space, up to $\log\log$ factors, when $p = 1$ (\Cref{thm:F_1}). Finally, one might wonder why along with \cite{PCVM2024feasibility} we only consider insertion-only streams rather than allow for turnstile updates. We address this by giving an $\widetilde{\Omega}(n)$ lower bound against $F_p$ moment estimation for $p = 0$ and $p \geq 1$ in turnstile streams (including strict turnstile streams), even for obtaining constant-factor approximations when $1-\alpha = O(1)$ (\Cref{thm:lb_turnstile}).
\begin{table}[]
    \small
    \centering
    \begin{tabular}{ |l|ll|ll|ll|l|} 
		\hline
		\textbf{$F_p$ } 
		&\multicolumn{2}{c|}{\textbf{Prior Work} } & \multicolumn{2}{c|}{\textbf{Our Work}} & \multicolumn{2}{c|}{\textbf{Lower Bound}} \\
		\hline
		$p = 0$					& $O\left(\frac{1}{\eps^2(1-\alpha) }\log n\right)$ &\hspace{-1em}\cite{PCVM2024feasibility} & \multicolumn{2}{c|}{\textbf{--}} & \cellcolor{blue!15}$\Omega\left(\frac{1}{\eps^2(1-\alpha) }\log n\right)$ & \cellcolor{blue!15}\hspace{-1em}(Thm. \ref{lem:lb_F0F1})\\

		$p = 1$ 						& $O\left(\frac{1}{\eps^2(1-\alpha) }\log^2 n\right)$ &\hspace{-1em}\cite{PCVM2024feasibility} & \cellcolor{blue!15}$\widetilde{O}\left(\frac{1}{\eps^2(1-\alpha) }\log n\right)$ & \cellcolor{blue!15}\hspace{-.8em} (Thm.~\ref{thm:fpopt1})& \cellcolor{blue!15}$\Omega\left(\frac{1}{\eps^2(1-\alpha) }\log n\right)$&  \cellcolor{blue!15}\hspace{-1.3em} (Thm. \ref{lem:lb_F0F1}) \\

		$p \in (0, 2)$ 		& \multicolumn{2}{c|}{\textbf{--}} &\cellcolor{blue!15}$\widetilde{O}\left(\frac{1}{\eps^2(1-\alpha) }\log n\right)$ &  \cellcolor{blue!15}\hspace{-.5em}(Thm. \ref{thm:fpopt1}) &\cellcolor{blue!15}$\Omega\left(\frac{1}{\eps^2(1-\alpha) }\log n\right)$ & \cellcolor{blue!15}\hspace{-1em}(Thm. \ref{lem:lb_F0F1}) \\

		$p = 2$ 		& \multicolumn{2}{c|}{\textbf{--}} &\cellcolor{blue!15}$\widetilde{O}\left(\frac{1}{\eps^2(1-\alpha) }\log^2 n\right)$ &  \cellcolor{blue!15}\hspace{-.5em}(Thm. \ref{thm:fpopt1}) &\cellcolor{blue!15}$\Omega\left(\frac{1}{\eps^2(1-\alpha) }\log n\right)$ & \cellcolor{blue!15}\hspace{-1em}(Thm. \ref{lem:lb_F0F1}) \\

		$p > 2$ 		& \multicolumn{2}{c|}{\textbf{--}} &\cellcolor{blue!15}$\widetilde{O}\left(\frac{1}{\eps^2(1 - \alpha)^{2/p}} \cdot n^{1 - 2/p} \right)$ &  \cellcolor{blue!15}\hspace{-.5em}(Thm. \ref{thm:fpopt}) &\cellcolor{blue!15}$\Omega\left(\frac{1}{\eps^2 (1-\alpha)^{2/p}}  n^{1 - 2/p}\right)$ & \cellcolor{blue!15}\hspace{-1em}(Thm. \ref{thm:lb_Fp}) \\
		\hline
	\end{tabular}
    
    \caption{Upper Bounds and Lower Bounds for $F_p$ estimation with forgets. $\widetilde{O}(\cdot)$ hides $\poly \log$ factors.}
    \label{tab:1}
\end{table}

\paragraph{Entropy Estimation with Forgets.} We extend our results for $F_p$ estimation to entropy estimation in the forget model. Formally, assuming that $|H(\ff) - H(\widetilde{\ff})| \leq \alpha H(\widetilde{\ff})$ and $F_1(\ff) \geq (1 - \alpha) F_1(\widetilde{\ff})$, we present a one-pass streaming algorithm that estimates the entropy of $\ff$ to within an additive error of $\eps$, using $\widetilde{O}\left(\frac{1}{\eps^2(1 - t\alpha)} \cdot \mathrm{poly} \log n \right)$ bits of space, where $t = 1 + O(1/\log n)$
(\Cref{thm:entropy}).

\paragraph{Heavy-hitters with General Operations.}
We develop an algorithm for estimating $F_p$ norms with $p \geq 2$ over data streams that support general update operations. Specifically, updates are of the form $(i, g)$, where $g \in \mathcal{G}$ is a contracting function applied as $f_i = g(f_i)$. A crucial component of our approach is the design of an $\ell_2$-heavy-hitters data structure capable of handling such updates efficiently in a streaming setting. This structure allows us to identify $\ell_2$-heavy hitters using $\widetilde{O}\left(\frac{1}{(1-\alpha)\varepsilon^2} \cdot \log n \right)$ bits of space (\Cref{thm:l2_hh}). Building on this as a black box, we further construct an algorithm to find $\ell_p$-heavy hitters, requiring $\widetilde{O}\left(\frac{1}{(1-\alpha)^{2/p} \varepsilon^2} n^{1 - 2/p}\right)$ bits of space (\Cref{thm:ell_p_heavy_hitters}). Importantly, our method not only identifies the indices of the heavy hitters but also approximates their values within an additive error of $\varepsilon \|\ff\|_2$.

\paragraph{$F_p$ Estimation with General Operations.}  
For $F_p$ estimation for $p > 0$ on a stream with general operations from $\mathcal{G}$, we design an algorithm that given $\eps \in (0,1)$ obtains a $(1 \pm \eps)$-approximation to $F_p$ with constant probability using $\widetilde{O}\left(\frac{1}{(1-\alpha)^{2/p}}\frac{1}{\eps^{2 + 4 / p}} n^{1 - 2/p} \right)$ bits of space for $p > 2$ and $\widetilde{O}\left(\frac{1}{(1-\alpha)^{2/p}}\frac{1}{\eps^{2 + 4 / p}} \cdot \poly \log n \right)$ bits of space for $0 < p \le 2$ (\Cref{thm:F_p}). 

\paragraph{Prefix-suffix Deletion Model.}
For the $F_1$ moment, we design an algorithm that, given parameters $\varepsilon, \delta \in (0,1)$, computes a $(1 \pm \varepsilon)$ approximation to $F_1$ with probability at least $1 - \delta$, using $\widetilde{O}\left(\frac{1}{(1-\alpha)\varepsilon^2} \log n \log\left(\frac{1}{\delta}\right)\right)$ bits of space (\Cref{thm:F1_psd_model}).

We also present an algorithm that, given $\varepsilon \in (0,1)$, estimates all coordinates of the frequency vector $\ff$ to within $\varepsilon \|\ff\|_2$ additive error using $\widetilde{O}\left(\frac{1}{(1-\alpha)\varepsilon^2} \cdot \log^2 n\right)$ bits of space (\Cref{thm:l2_hh_PSD_model}). This coordinate-wise estimator serves as a key subroutine in our final algorithm for approximating $F_p$ for any $p \geq 2$.
Specifically, for $\varepsilon \in (0,1)$ and $p \geq 2$, our algorithm outputs a $(1 \pm \varepsilon)$ approximation to $F_p$ using $\widetilde{O}\left(\frac{n^{1 - 2/p}}{(1 - \alpha)^{2/p} \varepsilon^{2 + 4/p}}\right)$ bits of space (\Cref{thm:F_p_PSD_model}).

\subsection{Our Techniques}
\paragraph{$\ell_p$ Sampling for $p \in (0, 2]$.} At a high level, we adopt a similar approach to that of~\cite{JW2021perfect}, which leverages the order statistics of exponential random variables. Specifically, define \( z_i = f_i / e_i^{1/p} \), where \( e_i \) are i.i.d.\ exponential random variables, and let \( D(1) = \arg\max_i |z_i| \). Then it holds that
\[
\Pr[D(1) = i] = \frac{|f_i|^p}{\|\ff\|_p^p}\;.
\]
Moreover, with probability \(\Omega(1)\) (or \(\Omega(1/\log n)\) when \(p = 2\)), we have \( |z_{D(1)}| = \Theta(1) \cdot \|\bz_{-D(1)}\|_2 \). The core idea in~\cite{JW2021perfect} is to generate multiple independent groups of exponential random variables and identify the maximal index whenever this event occurs. To achieve this,~\cite{JW2021perfect} employs a CountSketch-style data structure, which requires at least \(\Omega(\log^2 n)\) bits of space.

Recall that we are working in an insertion-only stream. We now turn our attention to another data structure, BPTree~\cite{BCINWW2017bptree}, which identifies \((\varepsilon, \ell_2)\)-heavy hitters using a more space-efficient approach, requiring only \(O(\varepsilon^{-1} \log n)\) bits of space. The hope is to use BPTree to identify the maximal index. However, this introduces a new challenge. To achieve the desired probabilistic guarantee, the exponential random variables need to exhibit full randomness. Since BPTree is not a linear sketch, we cannot directly apply the derandomization technique from~\cite{JW2021perfect}. Moreover, standard derandomization approaches would increase the \(\log n\) factor in the space complexity.

To overcome this barrier, we open the procedure of BPTree. At a high level, the key sub-routine of the BPTree is the following: assuming there is a single coordinate $f_i$ that constitutes an $O(1)$-fraction of the total mass of the frequency vector $\ff$, the goal is to identify this $O(1)$-heavy hitter,  To solve this problem, BPTree takes $O(\log n)$ rounds and in each round, it decides one bit of the heavy coordinate's identity. We instead give an alternative approach to solve the same heavy-hitter problem in the same $\widetilde{O}(\log n)$ space via inspiration from adaptive sparse recovery~\cite{IPW2011power}.  The key advantage of our method is that it requires only $O(\log\log n)$ rounds of linear sketches, compared to the $\log n$ rounds required by BPTree. This structural difference enables us to apply the derandomization technique from~\cite{JW2021perfect}, while incurring only an $O(\poly(\log\log n))$ space overhead.

In particular, consider a fixed $O(\log \log n)$ branching points. That is, we fix the $O(\log \log n)$ branching time $t_1, t_2, \dots, t_r$ and the corresponding subset $S_1, S_2, \dots, S_r$ it goes into. We refer to this fixed sequence as a {\em{path}}. The key observation here is that once these branching points are fixed, our algorithm will behave as a linear sketch. This allows us to apply the derandomization result from~\cite{JW2021perfect}
to show that
\[
\abs{\Pr(\texttt{path}) -\Pr'(\texttt{path})} \leq \frac{1}{n^{O(\log \log n)}}
\]
where $\Pr(\texttt{path})$ denotes the true probability of taking a path in the random oracle model and $\Pr'(\texttt{path})$ denotes the probability of taking a path after derandomizing.
 Note that there are at most $n^{O(\log \log n)}$ paths. Hence, we can take a union bound over all possible paths to obtain the desired derandomization guarantee.
 \paragraph{$F_p$ Estimation with Forgets ($0 < p \le 2$).} 
Let $\ff$ denote the frequency vector without the forget operations, and let $\bg$ denote the actual frequency vector after the forget operations. Our estimator is based on the $\ell_p$ sampling algorithm.
At a high level, let $j$ be an index sampled from the $\ell_p$ sampler, where
\[
\Pr[j = i] = \frac{|f_i|^p}{\|\ff\|_p^p}.
\]
Suppose we can construct an unbiased estimator $\hat{g}_j^p$ for $g_j^p$, and an unbiased estimator $\hat{\frac{1}{p_j}}$ for $\frac{1}{p_j}$, where $p_j = |f_j|^p / \|\ff\|_p^p$ is the sampling probability of index $j$. Then $\hat{\frac{1}{p_j}} \cdot \hat{g}_j^p $ serves as an unbiased estimator for $F_p(\bg) = \sum_{i=1}^n g_i^p$.
Moreover, if we can show that the variance of this estimator is bounded, we can reduce the overall variance by averaging over multiple independent samples.
However, several technical challenges remain:
\begin{itemize}
    \item How can we obtain an unbiased estimator of $g_j^p$? This is non-trivial because the stream allows forget operations, which may make it difficult to estimate $g_j$ accurately.
    \item How can we obtain an unbiased estimator of the inverse sampling probability $1/p_j$? Note that even if an estimator $h_j$ satisfies $\mathbf{E}[h_j] = f_j$, it does not follow that $\mathbf{E}[1/h_j] = 1/f_j$.
\end{itemize}

We address these challenges in the following sections.

\paragraph{Unbiased estimator of $g_j$.} Our estimator is motivated by the following crucial observation. Suppose that $g_j < (1 - \eps) f_j$. Then there must have been a forget operation that occurred after the frequency of item $j$ had reached at least $\eps f_j$ in $\ff$.
Since $j$ is a sampled coordinate from the $\ell_p$ sampler, it follows that $j$ is an $O(1)$-heavy hitter in the rescaled frequency vector $\bz$. Therefore, if we track the $\eps/10$-heavy hitters of $\bz$ and maintain exact counters for those items, we will be able to detect this forget operation. In this case, we can recover the exact value of $g_j$, since updates prior to the forget event can be ignored.
On the other hand, if $g_j \ge (1 - \eps) f_j$, then $f_j$ itself provides a valid $(1 \pm \eps)$-approximation to $g_j$, which suffices for our purposes.

However, within each sampler, maintaining the $\eps$-heavy hitter requires at least $\Omega(1/\eps^2)$ space. If we use $1/\eps^2$ independent samplers to reduce variance, the total space usage becomes $\Omega(1/\eps^4)$, which is not optimal for our setting. To address this space issue, we propose the following alternative estimator for $g_j$. For each sampler, we uniformly sample a random threshold $w \in (\eps, 1)$ and track the $w/10$-heavy hitters of the stream. Our final estimate for $g_j$ is defined as
\[
\mathbf{1}\left( \frac{g_j}{\hat{f}_j} \ge 1 - w \right) \cdot \hat{f}_j',
\]
where $\hat{f}_j$ and $\hat{f}_j'$ are two independent estimators of $f_j$.
The first question is whether we can compute this expression exactly. The key observation is the following:

\begin{itemize}
    \item If $g_j < (1 - w/3) f_j$, then—because we are tracking the $w/10$-heavy hitters—we will be able to recover the exact value of $g_j$.
    \item Otherwise, if $g_j \ge (1 - w/3) f_j$, then the indicator variable evaluates to $1$, and we simply return $\hat{f}_j'$.
\end{itemize}
Clearly, for the true value of $f_j$, we have
\[
\E\left[\mathbf{1}\left( \frac{g_j}{f_j} \ge 1 - w \right)\right] = \frac{g_j}{f_j (1 - \eps)},
\]
since $w$ is uniformly sampled from the interval $(\eps, 1)$. We now explain why the estimator remains (almost) unbiased when we use $\hat{f}_j$ in place of the true $f_j$.
When $g_j \ge (1 - w/3) f_j$, and $\hat{f}_j$ is a $(1 \pm O(w))$-approximation to $f_j$, then with high probability, the ratio $g_j / \hat{f}_j$ stays below $1$. Conditioning on this, we perform a careful probabilistic analysis and show that the estimator remains unbiased as long as
\[
\E\left[\frac{1}{\hat{f}_j}\right] = \frac{1}{f_j}.
\]
We will address how to ensure this property in a later section. For further technical details, we refer the reader to \Cref{sec:f_p2}.

\paragraph{Reducing to $1/\eps^2$ dependence.}
However, since each $w_i$ is uniformly sampled from $(\eps, 1)$, taking $1/\eps^2$ independent samples results in an expected space usage of
$\eps^{-2} \cdot \E[w^{-2}] = O\left(\eps^{-3}\right)$,
which is still sub-optimal.
To further reduce the space complexity, we observe that we obtain more samples than necessary in each subrange of $w_i$ values. This suggests a sub-sampling strategy within each level. Specifically, suppose we have $1/\eps^2$ independent $\ell_p$ samplers. Consider a fixed level $[\eps \cdot 2^\ell, \eps \cdot 2^{\ell+1}]$ for some $\ell \in \{0, 1, \dots, \log(1/\eps)\}$. The expected number of $w_i$ values falling into this level is roughly $N_\ell = O\left(\frac{2^\ell}{\eps}\right)$. We then randomly subsample $N_\ell' = O(4^\ell)$ of them and assign each corresponding estimator a weight of $N_\ell / N_\ell'$. This preserves unbiasedness of the overall estimator.
Although the variance may increase significantly at certain levels due to sub-sampling, we perform a careful analysis and show that the total variance across all levels increases by at most an $O(\log(1/\eps))$ factor.

\paragraph{Unbiased estimator of $1/f_j$.} The remaining task is to give an unbiased estimation of $1/f_j$ as well as the reverse sampling probability $\norm{\ff}_p^p / |f_j|^p$. Here we consider the Taylor expansion of $f(x) = 1/x$. In particular, suppose that $T$ is a $(1 \pm \frac{1}{2})$-approximation to $f_j$, we can expand $1/f_j$ as
\[
\frac{1}{f_j} = \sum_{i = 0}^\infty \frac{(-1)^{i} \cdot (f_j-T)^i}{T^{i+1}} =\frac{1}{T} - \frac{(f_j-T)^1}{T^2} + \frac{(f_j-T)^2}{T^3} - \ldots
\]
Since we only require a $(1 \pm \eps)$-approximation in expectation, and $T$ is a $(1 \pm \frac{1}{2})$ approximation to $f_j$, it suffices to truncate this series after the first $O(\log(1/\eps))$ terms. To estimate each term of the form $(f_j - T)^q$, we generate $q$ independent estimators $\hat{f}_j^{(1)}, \hat{f}_j^{(2)}, \dots, \hat{f}_j^{(q)}$ of $f_j$ and compute the product $\prod_{i = 1}^{q} (\hat{f_j^{(i)}} - T)$, which is an unbiased estimator for $(f_j - T)^q$ assuming each $\hat{f}_j^{(i)}$ is unbiased. This allows us to construct an (approximately) unbiased estimator for $1/f_j$ using the truncated Taylor expansion.

For the reverse sampling probability $\norm{\ff}_p^p / |f_j|^p$, we adopt a similar strategy to estimate $1/|f_j|^p$. For the numerator $\norm{f}_p^p$, we apply a standard technique based on $p$-stable distributions, using the geometric mean of several independent sketches. By multiplying these two components—the estimator for $\|\ff\|_p^p$ and the estimator for $1 / |f_j|^p$—we obtain an (approximately) unbiased estimator for the inverse sampling probability $\|\ff\|_p^p / |f_j|^p$.
\paragraph{$F_p$ Estimation with Forgets $(p > 2)$.}

We consider a similar algorithm to the one used for the case $0 < p \le 2$, but replace the $\ell_p$ sampling with $\ell_2$ sampling. By leveraging the standard relationship between the $\ell_p$ and $\ell_2$ norms, we show that it suffices to take  
$O\left(\frac{1}{\eps^2(1 - \alpha)^{2/p}} \cdot n^{1 - 2/p}\right)$ samples to achieve the desired accuracy.

\paragraph{$F_1$ Estimation with Forgets.}  We show that there exists a surprisingly simple algorithm for estimating $F_1$ in the $\alpha$-RFDS model. In the absence of forget requests, this task would be straightforward—we could simply maintain a count of the number of insertion operations. While this naïvely requires $O(\log m)$ space (where $m$ is the stream length), Morris counters can reduce this to an $O(\log \log m)$ dependence.

To handle forget requests, we estimate the number of insertion operations into the stream that are later erased by a forget request.  A random sample of insertion operations of size $O(\frac{1}{1-\alpha}\frac{1}{\eps^2})$ would suffice for this, and we could simply store the identities of the corresponding insertion indices using $O((\frac{1}{1-\alpha}\frac{1}{\eps^2})\log n)$ bits total and check if they are deleted after each operation that we sampled.

Since we cannot sample insertions in advance, we use reservoir sampling to construct the sample online. A technical challenge is that we do not know the exact stream length, as it is estimated using a Morris counter. This introduces slight noise into the sampling probabilities, which might be expected to accumulate over time. However, by carefully bounding the estimation error, we show that the impact on overall space remains negligible. We also note that one could alternatively use the reservoir sampling method of Gronemeier and Sauerhoff~\cite{GS2009applying}, though our analysis is simpler using the Morris+ counter~\cite{NY2022optimal}.

\paragraph{Heavy Hitters with General Operations.} 
Here we are allowing a group of more general functions including the forget operation. For a class of functions $\mathcal{G}$ which we will describe shortly, at each time $t$ in the stream, we allow an update of the normal addition $(i, +)$ or of the type $(i, g_t \in \mathcal{G})$ which performs $f_i = g_t(f_i)$. For ease of presentation, here we consider $\mathcal{G}$ as the class of contracting functions, or functions with $g : \mathbb{N} \to \mathbb{N}$ with the following conditions: $|g(x) - g(y)| \le |x - y|$.

We now consider the $\ell_2$ heavy hitter problem, which can be naturally extended to the $\ell_p$ heavy hitter problem for any $p \geq 2$. Let $\widetilde{\ff}$ denote the frequency vector excluding the operations in $\mathcal{G}$, and let $\ff$ be the true frequency vector that includes the operations in $\mathcal{G}$. The key observation is that any $\eps$-heavy hitter of $\ff$ is guaranteed to be an $\eps'$-heavy hitter of $\widetilde{\ff}$, where $\eps' = (1 - \alpha)^{1/2} \eps$.
Motivated by this, we track the $\eps'/10$-heavy hitters of $\widetilde{\ff}$ and maintain exact counters (initialized to zero) for these items. When a true heavy hitter of $\ff$ appears and is added to this list, its frequency may initially be underestimated by up to $\eps' \norm{\widetilde{\ff}}_2 = \eps \norm{\ff}_2$, due to the counter starting at zero. However, since the operations in the stream are either insertions or contractive transformations, the error in the estimate for this coordinate can only decrease or remain unchanged over time.

\paragraph{$F_p$ Estimation with General Operations.}
Our algorithms are inspired by the level-set arguments in the literature (see, e.g., Section 5 in~\cite{LWY2021exponentially}). Take $M$ to be an upper bound on the stream length. At a high level, we divide our vector's entries into level sets and let $S_j$ denote the set of coordinates $f_i$ where $f_i\in [\frac{M}{2^{j + 1}}, \frac{M}{2^j}]$ at the end of the stream, we then estimate the $s_j = \sum_{i \in S_j} |f_i|^p$ individually. Then the final estimation becomes
\[
s = \sum_i |f_i|^p = \sum_j \sum_{i \in S_j} |f_i|^p = \sum_j s_j \;.
\]
A crucial observation is we only need to estimate $s_j$ which takes up at least $\eps$-fraction of $F_p$. To get a good approximation to $s_j$, we can do the following. We sample the coordinates of the underlying vector with probability $2^{-i}$ for $i = 0, 1, \cdots, \log M$. That way, we can find the proper sampling rate where the coordinates in $S_j$ are heavy hitters in the subsampled stream. Therefore, we can use our heavy hitters structure to find the coordinates. By averaging and rescaling, we then can get a good approximation to $s_j$. 
   
\paragraph{Lower Bounds.}
Our lower bounds for $F_p$ estimation for $p \in [0, 2]$ are derived via a variant of the GapHamming communication problem, known as \emph{GapAndIndex}, which was also utilized by~\cite{PCVM2024feasibility}. In the standard GapAndIndex problem, Alice and Bob each hold $n$-bit strings and wish to estimate the number of indices where both bits are $1$, up to an additive error of $\sqrt{n}$. We consider a \emph{one-way} variant where Alice's vector is sparse—containing exactly one $1$ in each of $r$ blocks, each of size $n/r$—and the approximation must be within additive error $\sqrt{r}$\footnote{To facilitate the reduction, we actually work with a slightly modified version of this problem.}.
We show that solving this variant requires $\Omega(r \log(n/r))$ bits of communication. Setting $r = 1/\eps^2$ yields an instance that can be solved by a $(1 \pm \eps)$-approximation algorithm for $F_p$ in the $O(1)$-RFDS model. Specifically, Alice inserts items corresponding to the $1$-bits in her input, while Bob issues forget operations on the indices where his input has $1$s. The resulting $F_p$ moment accurately approximates the number of shared $1$-bits between Alice and Bob.

To extend our lower bound beyond the regime where $1-\alpha$ is constant, we take a direct sum of $1/(1-\alpha)$ hard instances with $1-\alpha$ constant.  After Alice sends her sketch to Bob, Bob can choose to delete all but one instance of his choice, so in fact he must be able to solve each one individually with good probability. By applying direct sum theorem, we then establish an $\Omega\left(\frac{1}{1-\alpha}\frac{1}{\eps^2}\log n\right)$ lower bound.

For our $F_p$ lower bounds with $p\geq 2$, we use an off-the-shelf communication lower bound for $F_p$ moment estimation.  This lower bound is also structured to admit a direct sum lower bound as above, for a sum of $\frac{1}{1-\alpha}$ instances. Applying this idea similarly as for $F_p, \ p \le 2$ yields our lower bound of $\Omega\left(\frac{1}{(1-\alpha)^{2/p}}\frac{1}{\eps^2}n^{1-2/p}\right).$

Finally, our lower bound for strict turnstile streams follows from a simple reduction from the well-studied set-disjointness communication problem, under the promise that the intersection size is either $0$ or $1$.  Interestingly, our reduction from set-disjointness results in a three-round protocol, so we use a hardness result for set-disjointness for multi-way protocols. 
\subsection{Related Work} \label{intro:RL}

There is a large body of research on estimating the frequency moments of data streams, and we do not aim to survey it exhaustively. Instead, we highlight key results for one-pass algorithms in arbitrarily ordered streams.
The foundational work of Alon, Matias, and Szegedy~\cite{AMS1999space} introduced a turnstile streaming algorithm for estimating $F_2$, using $O(\log n / \varepsilon^2)$ bits of space. A lower bound of $\Omega(\log n + 1/\varepsilon^2)$ was established by Woodruff~\cite{W2004optimal}, and this was later tightened to $\Omega(\log n / \varepsilon^2)$ by Braverman and Zamir~\cite{BZ2024optimality} for an insertion-only stream, who also resolved the space complexity for all $p \in (1,2]$, showing it to be $\Theta(\log n / \varepsilon^2)$.
For $p > 2$, Bar-Yossef et al.~\cite{BJKS2004information} and Chakrabarti, Khot, and Sun~\cite{CKS2003near} showed that estimating $F_p$ requires at least $\Omega(n^{1 - 2/p} / \varepsilon^{-2/p})$ space. The first algorithm to achieve optimal dependence on $n$ in this regime was given by Indyk and Woodruff~\cite{IW2005optimal}, using $\widetilde{O}(n^{1 - 2/p}) \cdot \text{poly}(\varepsilon^{-1})$ space. Later, Li and Woodruff~\cite{LW2013tight} essentially settled the space complexity at $\Theta(n^{1 - 2/p }\cdot\log n / \varepsilon^2)$.
In the case of $p = 0$, Woodruff~\cite{W2004optimal} proved a lower bound of $\Omega(\log n + 1/\varepsilon^2)$, which Kane, Nelson, and Woodruff~\cite{KNW2010optimal} matched with a tight algorithm in an insertion-only stream. For $p = 1$ in insertion-only streams, Nelson and Yu~\cite{NY2022optimal} showed the complexity is $\Theta(\log \log n + \log \varepsilon^{-1})$. For $p \in (0,1)$ in insertion-only streams, Jayaram and Woodruff~\cite{JW2023towards} provided a nearly tight algorithm with space complexity $\widetilde{O}(\log n + 1/\varepsilon^2)$.
The first algorithm for estimating $F_p$ for $0 < p < 2$ was proposed by Indyk~\cite{I2006stable}, using $p$-stable distributions to achieve a $(1 \pm \varepsilon)$-approximation with $O(\varepsilon^{-2} \log n)$ words of space. Kane, Nelson, and Woodruff~\cite{KNW2010exact} later improved this to $O(\varepsilon^{-2} \log n)$ \emph{bits}, which is optimal for the turnstile model.
Frequency moment estimation has also been studied in other settings, including random-order streams~\cite{BVWY2018revisiting}, multi-pass streaming~\cite{WZ2021separations}, and under differential privacy~\cite{EMMMVZ2023differentially,WPS2022differentially}.

There is also a long line of work on  $\ell_p$ sampling or sampling algorithms in a data stream \cite{CCD12, GLH07, CPW20}. For $p = 1$ in insertion-only streams, the reservoir sampling algorithm of Vitter~\cite{V1985random} solves the problem in $O(\log n)$ bits of space. For $p \in [0,2]$ (and in turnstile streams), Monemizadeh and Woodruff~\cite{MW2010pass} show that for approximate samplers, i.e. those that return an index $i$ with probability $(|f_i^p| / \| \ff \|_p^p)  \cdot (1 \pm \eps) + \poly(1/n)$ for $\eps \in (0,1)$, there is an algorithm using $\poly(\eps^{-1}, \log n)$ space that has probability of failure $\delta = 1/\poly(n)$. For $p \in [1,2]$, this was improved by Andoni, Krauthgamer, and Onak~\cite{AKO2010streaming} to $O(\eps^{-p}\log^3 n)$ bits. Jowhari, Sa\u{g}lam, and Tardos~\cite{JST2011tight} gave a sampler for $p \in (0,2) \setminus \{1\}$ using $O(\eps^{-\max(1,p)} \log^2 n)$ bits and a sampler for $p = 1$ using $O(\eps^{-1} \log(1/\eps)\log^2(n))$ bits. When we have $\eps = 0$, the samplers are called {\em{perfect}} samplers. Here, Frahling, Indyk, and Sohler~\cite{FIS2008sampling} gave a perfect $\ell_0$ sampler using $O(\log^2 n)$ bits. Jayaram and Woodruff~\cite{JW2021perfect} give perfect $\ell_p$ samplers for $p \in (0,2)$ using $O(\log^2 n (\log \log n)^2)$ bits and for $p = 2$ using $O(\log^3 n)$ bits of space. Woodruff, Xie, and Zhou~\cite{WXZ2025perfect} give a perfect $\ell_p$ sampler for $p > 2 $ using $O(n^{1-2/p} \cdot \poly \log n)$ bits of space. The best known lower bound is $O(\log(1/\delta) \log^2 n)$ for turnstile streams by Kapralov, Nelson, Pachocki, Wang, Woodruff, and Yahyazadeh~\cite{KNPWWY2017optimal}. Notably, the recent work~\cite{PW25} presents a perfect $G$-sampler for a class of functions $\mathcal{G}$, achieving optimal space complexity of $O(\log n)$. However, it is difficult to directly compare their results with ours. First, for the moment function, their class $\mathcal{G}$ only includes $G(x) = |x|^p$ for $p \in [0, 1]$. Second, their algorithm relies on a random oracle; removing this assumption through derandomization would likely incur an additional $O(\log n)$ overhead.

\subsection{Road Map}
In \Cref{sec:prelim} we present the necessary preliminaries, including key definitions and the main algorithmic tools used throughout the paper.  In \Cref{sec:sampler} we give our nearly-optimal $\ell_p$ sampler for $p \in (0,2]$. In \Cref{sec:forget_new}, we give our algorithms for $F_p$ estimation in the forget model. In \Cref{sec:extensions} we show how to incorporate general operations for $F_p$ estimation for $p \geq 2$, give our algorithms for the prefix/suffix deletion model, and present our corollary on estimating entropy. Finally in \Cref{sec:LB} we present our lower bounds. 

\section{Preliminaries}
\paragraph{Streaming Model.} \label{sec:prelim}
We start with a vector $\ff = (f_1, \ldots, f_n)$, where all elements are initially zero. Updates arrive one-by-one as a stream. In insertion-only streams, updates are of the form $(i, 1)$, indicating the operation $f_i = f_i + 1$. For more general update models, each update is of the form $(i, g)$, applying the operation $f_i = g(f_i)$ for some function $g$. For example, a forget request to coordinate $i$ corresponds to setting $f_i = 0$. Another example is an update dividing the coordinate by two, i.e., $f_i = f_i / 2$. We assume that all functions applied to coordinates have finite precision.

In streaming algorithms, the input is too large to be stored in memory, so the algorithm must process updates online and output an answer using space sublinear in the input size. The goal is generally to minimize this space usage. Let $m$ denote the length of the stream; we assume the algorithm is given $m$ (or an upper bound on $m$) at initialization. We also make the standard assumption that $m = \poly(n)$.

\paragraph{Frequency Moments.}

We take the following definition from \cite{AMS1999space}. Let $A = (a_1, \ldots, a_m)$ be a sequence of elements, where each $a_i$ is a member of $N = \{1,2,\ldots,n\}$. Let $v_i = |\{j:a_j = i \}|$ denote the number of occurrences of $i$ in the sequence $A$, and define, for each $p \geq 0$, 
$F_p = \sum_{i=1}^n v_i^p.$ 

\paragraph{Heavy Hitters.}
Given a vector $\bv \in \mathbb{R}^n$, we say that an index $i \in [n]$ is an $(\eps, \ell_p)$-heavy-hitter for $\bv$ if $|v_i| \geq \eps \norm{\bv}_p.$

The most fundamental heavy-hitters problem is that of identifying a set of size $O(1/\eps^2)$ containing the set of $\ell_2$ heavy hitters. An optimal solution was given in the seminal work of Charikar, Chen, and Farach-Colton~\cite{CCF2002finding} who used a CountSketch data structure to solve the heavy-hitters problem. Roughly, CountSketch works by partitioning the coordinates of $\bv$ into approximately $\Omega(1/\eps^2)$ buckets so that each heavy hitter is likely to end up in its own bucket. Simply summing up the items in a bucket would result in an additive approximation to each heavy hitter that depends on $\norm{\bv}_1.$ This is improved by first randomly flipping the signs of elements in each bucket before summing, resulting in a much better additive error of $\eps \norm{\bv}_2$ for each heavy hitter.
Importantly for us, CountSketch gives estimates of the heavy hitter values, along with their identities. More specifically, by taking $O(\log \frac{1}{\delta})$ independent CountSketches, we obtain an additive approximation of $\eps \norm{\bv}_2$ to a given coordinate $v_i$ with probability at least $1-\delta.$

\paragraph{Morris counters.}
In \Cref{sec:F1}, we are interested in keeping a running count up to $m$, but are unwilling to pay the naive $\log(m)$ bits. This can be addressed using a Morris Counter introduced by Morris~\cite{M1978counting}. The idea is to keep a counter $i$, but for each new item, to increment $i$ with probability $2^{-i}$. After $m$ items have been seen, the value of $2^i$ is a constant factor approximation to $m$ with good probability. Since $i$ is effectively keeping count of $\log m$, the counter only requires $O(\log\log m)$ bits. This basic idea can be improved to give a $1 \pm \eps$ approximation to the count with a given failure probability $\delta$. A tight bound of $O(\log\log m + \log\log \frac{1}{\delta} + \log \frac{1}{\eps})$ has been established for this problem in~\cite{NY2022optimal}, using their Morris+ algorithm, which is a variation on the classic Morris counter.

\section{$\ell_p$ Sampling for $p \in (0,2]$} \label{sec:sampler}
In this section, we will give the construction of our $\ell_p$ sampler in an insertion-only stream. We first consider the case $0 < p < 2$ and give an overview of the perfect $\ell_p$ sampler in~\cite{JW2021perfect}. Later, we will extend it to the case when $p = 2$ in~\Cref{sec:l2sampling}. At a high level, the algorithm utilizes the order statistics of the exponential random variables.

\begin{definition}[Exponential random variables]
We say $e$ is a exponential random variable with scale parameter $\lambda>0$, if the probability density function for $e$ is 
\[p(x)=\lambda e^{-\lambda x}.\]
In particular, we say $e$ is a standard exponential random variable if $\lambda=1$.
\end{definition}

\begin{fact}[Scaling of exponentials]
    Let $t$ be exponentially distributed with rate $\lambda$, and let $\alpha>0$. Then $\alpha t$ is exponentially distributed with rate $\lambda / \alpha$.
\end{fact}

\begin{definition}[Anti-rank]
Let $\left(t_1, \ldots, t_n\right)$ be independent exponential random variables. For $k=1,2, \ldots, n$, we define the $k$-th anti-rank $D(k) \in[n]$ of $\left(t_1, \ldots, t_n\right)$ to be the values $D(k)$ such that $t_{D(1)} \leq t_{D(2)} \leq \cdots \leq t_{D(n)}$.
\end{definition}

\begin{fact}[\cite{N2006order}]

 For any $i=1,2, \ldots, n$, we have
\[\operatorname{Pr}[D(1)=i]=\frac{\lambda_i}{\sum_{j=1}^n \lambda_j}.\]
\end{fact} 
\begin{corollary}
    \label{cor:exp}
    Let $\ff \in \mathbb{R}^n$ be a frequency vector. Let $\left(e_1, \ldots, e_n\right)$ be i.i.d. exponential random variables with rate $1$. Define $z_i=f_i / e_i^{1 / p}$ and let $(D(1), \ldots, D(n))$ be the anti-rank vector of the vector $\left(z_1^{-p} \cdots z_n^{-p}\right)$. Then we have
\[
\Pr[D(1)=i] = \frac{\left|f_i\right|^p}{\|\ff\|_p^p}\;.
\]
\end{corollary}

\Cref{cor:exp} implies that the $\ell_p$ sampler only needs to output the maximal index $i^\star$ of $\bz$. To do this, the algorithm in~\cite{JW2021perfect} needs some gap between $z_{D(1)}$ and $z_{D(2)}$ to detect the maximal index $i^\star$. In particular, it runs a statistical test at the end and outputs $\texttt{FAIL}$ if the test is not satisfied. To ensure the probability of failing the statistical test does not depend on $i^\star$, \cite{JW2021perfect} duplicates each coordinate of $\bf$ for $n^c$ times for some constant $c$ to remove all the heavy items. They show that after this step, the following holds. 

\begin{lemma}[\cite{JW2021perfect}]
    \label{lem:test}
    For constant $p \in (0, 2]$ and $\nu \ge n^{-c / 60}$, $\Pr[\neg \texttt{FAIL} \ | \ D(1)] = \Pr[\neg \texttt{FAIL}] \pm \Tilde{O}(\nu)$ for every possible $D(1) \in [n]$.
\end{lemma}
Throughout this section, we will assume the frequency vector $\bf$ is after the duplication and that \Cref{lem:test} therefore holds. We will also need the following fact.
\begin{lemma}[Proposition 1, \cite{JW2021perfect}]
\label{lem:prop1}
    For any $s = 2^j \le n c^{-2}$ for some $j \in \mathbb{N}$, we have $\sum_{i=4s}^{N} z_{D(i)}^2 = O(\|F\|_p^2 / s^{2/p-1})$ if $p \in (0,2)$ is a constant bounded below 2, and $\sum_{i=4s}^{N} z_{D(i)}^2 = O(\log(n)\|F\|_p^2)$ if $p=2$, with probability $1 - 3e^{-s}$.
\end{lemma}

\subsection{Warm-up: BPTree}
\label{sec:bptree}

In this section, we will first assume our algorithm has access to a random oracle. Then in the later section we shall give a way to derandomize it. Our $\ell_p$ sampler will be based on the following data structure.

\begin{theorem}(BPTree, \cite{BCINWW2017bptree}) \label{lem:bpTHH} 
For any $\eps \in (0, 1)$, there is $1$-pass insertion-only streaming algorithm \texttt{BPTree} that, with probability at least $1-\delta$, 
returns a set of $(\frac{\eps}{2}, \ell_2)$ heavy hitters containing every $(\eps, \ell_2)$-heavy hitter. The algorithm uses $O(\eps^{-2} \log n\log \frac{1}{\eps \delta})$ bits of space
and has $O(\log \frac{1}{\eps \delta})$ update time and $O(\eps^{-2} \log \frac{1}{\eps \delta})$ retrieval time. 
\end{theorem}

Our algorithm is as follows. 
\begin{enumerate}
    \item Let $\bz$ be the vector defined in \Cref{cor:exp}.
    \item Using BPTree (\Cref{lem:bpTHH}), keep track of $\bz$ during the stream. In parallel, we use a CountSketch with $O(1)$ buckets and an AMS sketch (\cite{AMS1999space}) for $\ell_2$ norm to keep track of $\bz$.
    \item At the end of the stream, let the $\calS$ be the output set of $O(1)$-heavy hitters from the BPTree. For every $i \in \calS$, let $v_i$ be the estimation of $z_i$ using the Count-Sketch. Let $\Tilde{F}_2$ be an estimation of $\norm{\bz}_2^2$ from the AMS sketch. 

    \item Let $v_{D(1)}$ and $v_{D(2)}$ be the first and second largest value in $v_i \ (i \in \calS)$. If $v_{D(1)} - v_{D(2)} \ge c_1 \cdot \widetilde{F}_2$ for some constant $c_1$, output $\argmax_{i \in s} {v_i}$. 
     
\end{enumerate}

At a high level, we will run $O(\log\log n + \log(1/\eps))$ independent trials of the above subroutine and do the statistical test at the end. Then we output the maximal index $i$ from an arbitrary trial where the statistical test succeeds. If none of the trials succeed, the algorithm outputs $\texttt{FAIL}$.

\begin{theorem}
    \label{thm:sampler}
    For any constant $p \in (0, 2)$, there is a one-pass streaming algorithm that runs in space $\widetilde{O}(\log n \log(1/\eps))$, and outputs an index $i$ such that with probability at least $1 - \mathrm{poly}(\eps / \log n)$ we have for every $j \in [n]$,
    \[\operatorname{Pr}[i = j]=\frac{\left|f_j\right|^p}{\|\ff\|_p^p} \pm \frac{1}{\poly{(n)}} \;. \]
\end{theorem}

\begin{proof}
    Condition on the CountSketch for $v_{D(1)}$ and $v_{D(2)}$, and AMS sketch success for a constant approximation in all of the independent trials by taking a union bound. Then, consider an arbitrary trial where the statistical test succeeds. From the condition we have that using the CountSketch and AMS sketch we can find the maximal index $i^\star$ in $\calS$ that returned by the BPTree. On the other hand, from \Cref{lem:prop1} we have that with probability $\Omega(1)$ we have $z_{D(1)}^2 > 0.6 F_2$, this means that if we run $O(\log\log n + \log(1/\eps))$ independent trials, with probability at least $1 - \poly{(\eps / \log n)}$ one of the trials will pass the statistical test, and the output of this trial is what we need.

    Now we analyze the space complexity of our algorithm. For each independent trial, we set the accuracy parameter $\eps' = O(1)$ and error probability $\delta = \poly{(\eps / \log n)}$ for the CountSketch, AMS sketch and BPtree. This means the space we need for each trial is $\widetilde{O}(\log n \log(1/\eps))$ and the total amount of space is $\widetilde{O}(\log n \log(1/\eps))$.
\end{proof}

\subsection{Heavy-hitters via adaptive sparse recovery}
The difficulty with derandomizing the above $\ell_p$ sampling algorithm without introducing additional $\log n$ factors is that the BPTree is not a linear sketch. This means we cannot apply the derandomization technique from~\cite{JW2021perfect}. At a high level, the key sub-routine of the BPTree is the following: given a $2$-approximation of $F_2(\ff)$, find the single $O(1)$-heavy hitter of $\ff$. To solve this problem, BPTree takes $O(\log n)$ rounds and in each round, it decides one bit of the heavy coordinate.
In this section, we will give another way to solve this heavy-hitter problem in just $\widetilde{O}(\log n)$ space via inspiration from~\cite{IPW2011power}.  Compared to BPTree, the advantage of the algorithm here is that we only require $\log\log n$ rounds of linear sketches rather than $\log n$ for BPTree, which enables us to apply the derandomization technique from~\cite{JW2021perfect} while only blowing up our space by $O(\poly(\log \log n))$ factors. 

We recall the following lemma from \cite{IPW2011power} to show correctness of their sparse recovery algorithm.

\begin{lemma} (Lemma 3.2, \cite{IPW2011power})
\label{lem:one_round_of_adaptive_sparse_recovery}
Suppose that there exists $j$ with
$\abs{x_j} \geq C\frac{B^2}{\delta^2} \norm{\bx_{S\setminus \{j\}}}_2$ for a constant $C$ and parameters $B$ and $\delta.$ Then with two non-adaptive linear measurements, with probability $1-\delta$, we can find a set $S\subseteq n$ such that $|S| \le 1 + n/B^2$ and 
\[
\norm{\bx_{S\setminus \{j\}}}_2 \leq \frac{1}{B}\norm{\bx_{[n]\setminus\{j\}}}_2.
\]
\end{lemma}

First, we assume that we know $F_2(\ff)$ within a factor of two. Let $\hat{F}$ be that estimate of $F_2(\ff)$.
We will make use of the following $F_2$ tracker, and mark the times $t_i$ at which the tracker's $F_2$ estimate becomes at least $i\hat{F}/(\log\log n).$  

\begin{lemma}[\cite{BCINWW2017bptree}]
There is an algorithm that estimates $F_2$ at all times $i$ to within a constant factor with high constant probability, using $O(\log n (\log\log n + \log\frac{1}{\delta}))$ space.
\end{lemma}

We next note that without loss of generality we can further assume the heavy hitter $f_i$ satisfies $|f_i| \geq (1 - 1/\log \log n) \norm{\ff_{[n]\setminus \{i\}}}_2$, as we can randomly divide the universe into $10(\log \log n)^2$ groups and run our procedure in each groups. By Markov's inequality we have with probability at least  $1 - 1/\log \log n$, $f_i$ will be $(1 - 1 / \log \log n)$-heavy in its group. This will increase the space by only a $(\log \log n)^2$ factor.

\begin{lemma}
\label{lem:compress_sensing}
Suppose that some universe item $j$ satisfies
$|f_j| \geq (1 - 1/\log \log n) \norm{\ff_{[n]\setminus \{j\}}}_2$.  Then there is an algorithm to recover $j$ with probability larger than $1/2 $, using $\widetilde{O}(\log n )$ space.  Moreover this algorithm runs in only  $O(\log \log n)$ rounds and uses only $O(1)$ row linear sketches in each round.
\end{lemma}

\begin{proof}
We consider a similar strategy to that of~\cite{IPW2011power}. Let $B_0 = 2$ and $B_i = B_{i - 1}^{3/2}$ for $i \ge 1$ and define $\delta_i = 2^{-i} / 4$ for $i = 0$. Start with $S_0 = [n]$. Consider the stream of updates during time $t_{i}$ and $t_{i + 1}$, from the condition we have $f_j$ will be $(1 - 1/C)$-heavy during the stream between $t_i$ and $t_{i + 1}$ for some sufficiently large constant $C$. Applying \Cref{lem:one_round_of_adaptive_sparse_recovery} with parameters $B = 4 B_i$ and $\delta = \delta_i$, we can identify a subset $S_{i + 1} \subset S_i$ with $j \in S_{i + 1}$, ending with $i = r$ where $r = O(\log \log n)$ so $B_r \ge n$.

Similarly to~\cite{IPW2011power}, we prove by induction that \Cref{lem:one_round_of_adaptive_sparse_recovery} applies at the $i$-th iteration. 
We choose $C$ to match the base case.
For the inductive step, suppose $ \|x_{S \setminus \{j\}}\|_2 \leq |x_j| / (C' 16 \frac{B_i^2}{\delta_i^2})$. Then by \Cref{lem:one_round_of_adaptive_sparse_recovery},
\[
\|\bx_{S \setminus \{j\}}\|_2 \leq |x_j| / (C' 64 \frac{B_i^3}{\delta_i^2}) = |x_j| / (C' 16 \frac{B_{i+1}^2}{\delta_{i+1}^2}). 
\] 
This means that the lemma applies in the next iteration as well. Note that after $r$-th iteration we have $S_r < 2$, which means that we can identify $j \in S_r$. Taking a union bound, the overall failure probability is at most $\sum_i \delta_i < 1 / 2$.
\end{proof}

\subsection{Derandomization}
We next derandomize the $\ell_p$ sampler. We can see from~\Cref{lem:compress_sensing} that the procedure has $O(\log \log n)$ rounds or branching points. For some fixed $O(\log \log n)$ branching points (here we fix the $O(\log \log n)$ branching time $t_1, t_2, \dots, t_r$ and the subset $S_1, S_2, \dots, S_r$ it goes into), let us call this a path. Note that there are at most $n^{O(\log \log n)}$ paths. 

The main thing we wish to derandomize is the exponential re-scalings. Once we derandomize these scalings, it is possible that the probability of taking a path will differ than the probability of taking that path in the random oracle model. Therefore, we aim to also show that 
\[
\abs{\Pr(\texttt{path}) -\Pr'(\texttt{path})} \leq \frac{1}{n^{O(\log \log n)}}
\]
where $\Pr(\texttt{path})$ denotes the true probability of taking a path in the random oracle model and $\Pr'(\texttt{path})$ denotes the probability of taking a path after derandomizing. If we show this, then we are done since we can take a union bound over all the paths, giving us total variation distance $1/n^{O(\log \log n)}$. 

Note that the sketching matrix in~\Cref{lem:compress_sensing} only needs $O(1)$-wise independence, so the only part we need to derandomize is the exponential random variables part. This means that we can fix the value of the sparse recovery part and consider one fixed path. The advantage here is after this step, we can treat the sparse recovery part in~\Cref{lem:compress_sensing} as a linear sketching with $O(\log \log n)$ rows.
We use the following result from \cite{JW2021perfect}. 

\begin{lemma}(\cite{JW2021perfect}, Theorem 5)
Let \texttt{ALG} be any steaming algorithm which, on stream vector $\ff \in \{-M, \ldots, M\}^n $ and fixed matrix $\bX \in \R^{t \times nk}$ with entries contained within $\{-M, \ldots, M\}$, for some $M = \poly(n)$, outputs a value that only depends on sketches $\bA \cdot \ff$ and $\bX \cdot \texttt{vec}(\bA)$. Assume that the entries of the random matrix $\bA \in \R^{k \times n}$ are i.i.d. and can be sampled using $O(\log n)$ bits. Let $\sigma: \R^k \times \R^t \to \{0,1\} $ be any tester which measures the success of \texttt{ALG}, namely $\sigma(\bA \ff, \bX \cdot \texttt{vec}(A)) = 1$ whenever \texttt{ALG} succeeds. Fix any constant $c \geq 1$. Then \texttt{ALG} can be implemented using a random matrix $\bA'$ using random seed of length $O((k+t)\log n(\log\log n)^2)$, such that:  
\[
  \abs{\Pr[\sigma(\bA \ff, \bX \cdot \texttt{vec}(\bA)) = 1] - \Pr[\sigma(\bA' \ff, \bX \cdot \texttt{vec}(\bA'))=1]} < n^{-c(k + t)}. 
\]
\end{lemma}

Here, we take $\bA$ to be an $n$-dimensional row vector of i.i.d. exponential random variables.
, and $\bX = \bS \texttt{diag}(\ff) $ to be a $O(\log \log n) \times n$ fixed matrix where $\bS$ is the sketching matrix from \Cref{lem:compress_sensing}. 
Therefore, we have that for a fixed path, the probability difference $\abs{\Pr(\texttt{path}) -\Pr'(\texttt{path})} $ is at most $n^{-c \log \log n}$. This is sufficient for us as then we can take a union bound over all $n^{O(\log \log n)}$ paths.

\subsection{Extending to $\ell_2$ Sampling} 
\label{sec:l2sampling}
We next consider the case when $p = 2$. By Lemma~\ref{lem:prop1}, we have that with high constant probability $\norm{z_{-D(1)}}_2^2= O(\log n) \cdot F_2(\bz)$. Hence, with probability $\Omega\left(\frac{1}{\log n}\right)$ we have $|Z_{D(1)}|^2 > 20 \norm{z_{-D(1)}}_2^2$. This means that we will need to run the sub-routine in~\Cref{sec:bptree} $\widetilde{O}(\log(n) \log(1/\eps ))$ times.
\begin{corollary}
    There is a one-pass streaming algorithm that runs in space $\widetilde{O}(\log^2 n \log(1/\eps))$, and outputs an index $i$ such that with probability at least $1-\poly(\eps / \log n)$ we have for every $j \in [n]$,
    \[\operatorname{Pr}[i = j]=\frac{\left|f_j\right|^2}{\|\ff\|_2^2} \pm \frac{1}{\poly{(n)}} \;. \]
\end{corollary}

\subsection{Continuous $\ell_p$ Sampler}
In some scenarios, providing a sample index $i$ at the end of the stream is not enough. 
Instead, we may require the sampler to give a sample index at each time step during the stream. We next give a continuous $\ell_p$ sampler which is based on the $\ell_p$ sampler we presented in \Cref{thm:sampler}.

\begin{lemma}[Continuous $\ell_p$ sampler]
    \label{lem:l_p_sampler} 
    Consider a stream of length $m$ (and therefore with $m$ time steps). Let $F_{p, [0,t]}$ be the $F_p$ moment of $\ff$ after the first $t$ time steps, and let $f_{i, [0,t]}$ be the frequency of coordinate $i$ after the first $t$ time steps. 
    There is an algorithm $\mathcal{A}$ that produces a series of $m$ outputs $z_1, \ldots, z_m$ such that each $z_j \in \{1, \ldots, n\}$, one at each time step with the following properties:
    \begin{itemize}
        \item $\Pr(z_t = i) = \frac{f^p_{i, [0,t]}}{F_{p,[0,t]}} \pm \frac{1}{\poly(n)}$.
        \item Only $O(\log^{c(p)} n \cdot \log\log n\cdot \log(1/\eps))$ of the $z_j$'s are unique. 
        \item $\mathcal{A}$ uses $O(\log^{c(p)} n \cdot \polylog\left(1/\eps\right) \cdot \poly(\log \log n))$ bits of space.
        \item $\mathcal{A}$ succeeds with probability at least $1-\poly{(\eps / \log n)}$. 
    \end{itemize}
    Here $c(p) = 1$ from $0 < p < 2$ and $c(p) = 2$ for $p = 2$.
\end{lemma}
\begin{proof}

We first consider the case when $p \in (0, 2)$. Similarly to what we did in~\Cref{sec:bptree}, we run $O(\log(1/\eps) \cdot \log \log n)$ independent trials of the sub-routines and in each subroutine we use the BPTree to keep track of the vector $\bz$ scaled by the exponential random variables. Starting from $t = 1$, suppose that if we have a sample $z_{t - 1}$ at time $t - 1$ which corresponds to a particular trial of the exponential variables. Then we will set $z_t = z_{t - 1}$ if the statistical test for this specific trial still be satisfied at the time $t$. Otherwise, we will look at the other trials of the sub-routine and output $z_t$ from an arbitrary sub-routine where the statistical test passes.

From the correctness of~\Cref{thm:sampler} we get the correctness of properties (1) and (3). To bound the number of unique $z_i's$, note that every time when we change the value of $z_t$, the statistical test must have failed again for the corresponding trials. This implies the $F_2$ moment of the rescaled vectors must increase by a constant factor during this period of time. Since we have only $O(\log(1/\eps) \cdot \log \log(n))$ number of trials, we can get the number of the time $z_i$ changes is at most $O(\log n \cdot \log(1/\eps) \cdot \log \log(n))$. Finally, we consider the overall success probability. The only thing we need is when $z_i$ changes, at least one of the trials of the sub-routines passes the statistical test. Since for each trial, the success probability is $\Omega(1)$, from the fact that $z_i$ can only changes $O(\log n \cdot \log(1/\eps) \cdot \log \log(n))$ time and we have $O(\log (1/\eps) \cdot \log \log(n))$ trials of the sub-routines we have the overall success probability is at least $1 - \poly(\eps / \log n )$.

We next consider the case for $p = 2$. At the time, we run $\tilde{O}(\log n \log(1/\delta))$ independent trials of sub-routines, the remain argument still goes through.
\end{proof}

\subsection{Estimating the Value and Sampling Probability}

As mentioned, one application of our $\ell_p$ sampler is to estimate the $F_p$ moment. In this case in addition to an index $i$, we may also need an estimation of $f_i$ and the sampling probability $p_i = \frac{|f_i|^p}{\|\ff\|_p^p}$ (in fact, the inverse sampling probability).
We will next first show that we can get good enough estimators of the values of the coordinates that are sampled by the $\ell_p$ sampler. 

\begin{lemma}\label{lem:unbiased_est_of_fx}
    For a sampled index $i$ by an $\ell_p$ sampler from \Cref{thm:sampler}, we can get an estimator $\hat{f}_i$ to $f_i$ such that $\E[\hat{f}_i] = f_i $ and 
    $\E[\hat{f}_i^{2}] = O(f_i^{2})$. The space requirement is $\widetilde{O}(\log n)$ bits. 
\end{lemma}
\begin{proof}
    Let $\bz$ be the vector $\ff$ recaled by the exponential random variables $u_i^{1/p}$. We use a CountSketch with a constant number of buckets to estimate the frequency of the sampled index $z_i$ (this can be derandomized from the same procedure in~\cite{JW2021perfect}). By the property of CountSketch, we have $\E[\hat{z}_i] = z_i$. To bound the variance, 
    note that we have that $|z_i|^2 \geq \Theta(1) \| \bz \|_2^2$ since $i$ is the sampled coordinate. So, we have $\E[\hat{z}^2_{i}] = O\left(\norm{\bz}_2^2\right) = O\left(z^2_{i}\right)$. Multiplying by $u_i^{1/p}$ in both side we get the desired result.
\end{proof}

We now consider the estimation of $|f_i|^p$. For $p = 2$, we can simply run two independent estimator of $f_i$ and take their product. For other $p \in (0, 2)$ we will use the following Taylor's series estimator.
\begin{lemma} \label{lem:TSXP}
    Suppose $X$ is a fixed value in $[(1-\alpha \cdot T, (1+\alpha) \cdot T]$, and suppose that we have access to $Y_1, \ldots, Y_N$ where the $Y_j's$ are independent with $\E[Y_j] = X$ for some constant $C$ and $\Var[Y_j] \leq T^2 / 4$. 

    For $\alpha < \frac12$, we can construct an estimator $\hat{Q}$ of $X^p$ for $p \in [0,2)$ with $\abs{\E[\hat{Q}] - X^p} \leq \eps^{10} T^p$ and $\Var[\hat{Q}] = O(T^{2p}\log(1/\eps))$ for $N = O(\log^2 \frac{1}{\eps})$ for $\eps \in (0,1)$.
\end{lemma}
\begin{proof}
Consider the Taylor series expansion for $f(x) = x^p$ for $p \in [0,2)$ centered at $T$. 
For $p = 0$ we have $f(x) = x^0 = 1$. For $p = 1$ we have $f(x) = x = T + (x-T)$. For $p \in [0,2)$ but not $0$ or $1$, we have 
\[
    x^p = T^p + pT^{p-1} (x-T) + \frac{p(p-1)}{2!} T^{p-2}(x-T)^2 + \frac{p(p-1)(p-2)}{3!}T^{p-3}(x-T)^3 + \ldots
\]
valid for $x \in (0, 2T)$. 

Define $P_k(x)$ to be the associated degree $k$ Taylor polynomial. So, we have 
\[
P_k(x) = T^p \sum_{i=0}^k \binom{p}{i} \paren{\frac{x-T}{T}}^i.
\]

Recall that when truncating a Taylor series at degree $k$ for some function $f(x)$, the error is 
\[
\abs{P_k(x) - f(x)} = \abs{\frac{f^{k+1}(\xi)}{(k+1)!} \cdot (x-T)^{k+1}}
\]
where $\xi$ is some fixed point between $T$ and $x$, and $f^{k+1}(\xi)$ denotes the value of the $(k+1)$-th derivative of function $f$ at $\xi$. 

For $f(x) =x^p$, we have that
\[
f^{k+1}(\xi) = p(p-1)(p-2)\ldots(p-k) \cdot \xi^{p-k-1}.
\]
So, for $X \in [(1-\alpha) \cdot T, (1+\alpha) \cdot T]$, we have error 
\begin{align*}
\abs{P_k(X) - X^p}  &=  \abs{\frac{ p(p-1)(p-2)\ldots (p-k) \cdot \xi^{p-k-1} }{ (k+1)!} \cdot (X -T)^{k+1}}\\
&\leq \frac{(k+1)! \cdot \xi^{p-k-1}}{(k+1)!} \cdot \abs{(X-T)^{k+1}}
= \frac{\abs{(X-T)^{k+1}}}{\xi^{k+1-p}} 
\\ &\leq \frac{(\alpha T)^{k+1}}{((1-\alpha) \cdot T)^{k+1-p}} = \frac{\alpha^{k+1}}{(1-\alpha)^{k+1-p}} \cdot T^p. \tag{since $X \in [(1-\alpha) \cdot T, (1+\alpha) \cdot T]$}
\end{align*}

By the lemma statement, we have access to $Y_1, \ldots, Y_N$ which are independent unbiased estimators of $X$ and therefore to $X-T$ after shifting by $T$. For each $Y_j$ for $j \in [N]$ we have 
\begin{align*}
\E[(Y_j - T)^2] & \leq \E[2(Y_j-X)^2 + 2(X-T)^2] = 2\Var[Y_j] + 2(X-T)^2 \tag{since $(A+B)^2 \leq 2A^2 + 2B^2$ by AM-GM} \\
& \leq \frac{T^2}{2} + 2(\alpha T)^2 \leq T^2. \tag{since $\Var[Y_j] \leq T^2/4$ and $\alpha \leq 1/2$}
\end{align*}

Let us take $\ell$ of the independent estimators $Y_j - T$ and multiply them. Call this estimator $Y^\ell$. Since we have that the estimators $Y_j - T$ for $j \in [N]$ are independent, we have that $\E[Y^\ell] = (X - T)^\ell$. Furthermore, we have that 
\[\E[(Y^{\ell})^2]  = \E[((Y_1 - T)(Y_2 - T)\ldots (Y_\ell - T))^2] \leq (T^2)^\ell.
\]

By simply rescaling, we get an estimator $Z_\ell$ such that \[\E[Z_\ell] = \binom{p}{\ell}\frac{(X-T)^\ell}{T^{\ell-p}}, \E[Z_\ell^2] \leq \binom{p}{\ell}^2\frac{(T^2)^\ell}{T^{2(\ell-p)}} \leq 4 \cdot T^{2p}.
\]

Let us set our estimator $\hat{Q} = \sum_{i=1}^k Z_i$. By using our upperbound on $\abs{P_k(X) - X^p}$ above, we have 
\[
\abs{\E[\hat{Q}] - X^p} = \abs{\E[Z_1 + \ldots + Z_k] - X^p} \leq \frac{\alpha^{k+1}}{(1-\alpha)^{k+1-p}} \cdot T^p.
\]

We also have 
\[
\Var[\hat{Q}] = \Var[Z_1 + \ldots + Z_k] = \Var[Z_1] + \ldots + \Var[Z_k] \leq \E(Z_1^2) + \ldots + \E(Z_k^2)
\leq 4kT^{2p}.
\]
The result follows upon setting $k = \Theta(\log \frac{1}{\eps}).$
\end{proof}

By running $O(\log^2(1/\eps))$ independent copies of the estimator from~\Cref{lem:unbiased_est_of_fx}  and plugging this into~\Cref{lem:TSXP}, we get the following lemma. 

\begin{lemma}\label{lem:unbiased_est_of_fx_p}
    For a sampled index $i$ by an $\ell_p$ sampler from \Cref{thm:sampler}, we can get an estimator $v_i$ to $f_i^p$ such that $\E[v_i] = (1 \pm \eps^{10})|f_i|^p $ and 
    $\E[v_i^{2}] = O(|f_i|^{2p})$. The space requirement is $\widetilde{O}(\log n \log^2(1/\eps))$ bits. 
\end{lemma}

We next consider the estimation of the (reverse) sampling probability $1/p_i$. Recall that the sampling probability for an index $i$ is $f_i^p / \| \ff \|_p^p$. Here we show how to estimate the reverse sampling probability which will be used in our downstream applications. 

We first show how to get a rough estimate of $F_p$ (or $\| \ff \|_p^p)$ for \emph{constant} $p \in (0,2)$, which is part of the inverse sampling probability. To do this, we will use the geometric mean estimator and $p$-stable random variables. We first give the definition of $p$-stable random variables. 

\begin{definition}(Zolotarev~\cite{Z1986one}) \label{def:pstable}
    For $0 < p < 2$, there exists a probability distribution $\mathcal{D}_p$ called the $p$-stable distribution with $\E[e^{itZ}] = e^{-|t|^p}$ for $Z \sim \mathcal{D}_p.$ For any $n$ and vector $\bx \in \mathbb{R}^n$, if $Z_1, \ldots, Z_n \sim \mathcal{D}_p$ are independent, then $\sum_{j=1}^n Z_j \bx_j \sim \| \bx \|_p Z$ for $Z \sim \calD_p$. 
\end{definition}

The following fact will also be useful. 

\begin{lemma}(Nolan~\cite{N2020univariate}, Theorem 1.2) \label{lem:pstable}
    For fixed $0 < p < 2$, the probability density function of the $p$-stable distribution is $\Theta(|x|^{-p-1})$ for large $x$. 
\end{lemma}

When considering one $p$-stable random variable, they have undefined expectation and infinite variance. However, we will show that by using a geometric mean we can control the expectation and variance. We take $\bS$ to be a matrix of i.i.d. $p$-stable random variables with $k = \frac{2}{p-\eps'}$ where $\eps'$ is a small constant between $0$ and $1$. 
We consider $F$ which we call the {\it geometric mean estimator} which is based on the same sketch $\bS\bx $ (for a fixed $\bx \in \mathbb{R}^n$) with $k$ rows. 
In particular, we have 
\[F = |(\bS \bx)_a \cdot (\bS \bx)_b \cdot  (\bS \bx)_c \ldots |^{\frac{p - \eps'}{2}}.
\]

\begin{lemma} \label{lem:pstableest}
    $\E[F] = C \cdot \| \bx \|_p$ where $C > 0$ is a constant that does not depend on $\bx$. $\Var[F] = O(\| \bx \|_p^2)$. 
\end{lemma}
\begin{proof}
We have $(\bS \bx)_a = \|\bx \|_p  \cdot P_1$, $(\bS \bx)_b = \|\bx \|_p  \cdot P_2$, $(\bS \bx)_c = \|\bx \|_p  \cdot P_3$, and so on where $P_1, P_2, P_3, \ldots $ are independent $p$-stable random variables by \Cref{def:pstable}. Therefore we have that 
\[
\E[F_i] = \|\bx \|_p \cdot \E[|P_1 P_2 P_3 \ldots|^{\frac{p-\eps'}{2}}]. 
\]
We now show that $\E[|P_1 P_2 P_3 \ldots|^{\frac{p-\eps'}{2}}]$ is a finite scalar. Recall that $P_1, P_2, P_3,$ and so on are independent. Therefore we have 
\[
\E[|P_1 P_2 P_3 \ldots|^{\frac{p-\eps'}{2}}] = \int_{\frac{1}{\frac{p-\eps'}{2}}} C \cdot  \frac{|z_1 z_2 z_2 \ldots |^{\frac{p-\eps'}{2}}}{(z_1 z_2 z_2 \ldots)^{p + 1}} dz_1 dz_2 dz_3 \ldots 
\]
by \Cref{lem:pstable} which converges to a constant. The variance is similar. We have that 
\[
\Var[F_i] = \| \bx \|_p^2 \cdot \Var[|P_2 P_2 P_3 \ldots|^{\frac{p-\eps '}{2}}].
\]
By definition we have $\Var[|P_2 P_2 P_3 \ldots|^{\frac{p-\eps '}{2}}] \leq \E[|P_1 P_2 P_3 \ldots|^{p-\eps'}]$ and so we again get that this converges to a constant. 
    \qedhere
\end{proof}
We remark that the geometric mean estimator which we use in \Cref{lem:pstableest} to estimate the inverse sampling probability of a coordinate by our $\ell_p$ sampler was derandomized by \cite{KNW2011fast} (Theorem 12) with only an additive logarithmic factor. 

So, we can use \Cref{lem:pstableest} to get a unbiased (after dividing by $C$) estimator for $\| \ff \|_p$ with variance $O(\| \ff \|_p^2)$. 
To turn this into an estimator for $\| \ff \|_p^p$, we again run $\log^2(1/\eps)$ independent copies and plug into the Taylor's estimator in~\Cref{lem:TSXP}. 

To finish estimating the inverse sampling probability, we also need to estimate $1/f_i^p$. We can get an estimator for $f_i^p$ using \Cref{lem:unbiased_est_of_fx} and \Cref{lem:TSXP}. Now, we show how to estimate $1/f_i^p$ from this. 
The following lemma gives the precise guarantee that we will apply.

\begin{lemma} \label{lem:taylor}
Suppose that $X$ is a fixed value in $[(1-\alpha) \cdot T, (1+\alpha) \cdot T],$ and suppose that we have access to $Y_1, \ldots, Y_N$ where the $Y_j$'s are independent with $\E[Y_j] = X$ and $\Var[Y_j] \leq T^2/4.$

For $\alpha < \frac{1}{2}$, we can construct an estimator $\hat{Q}$ of $\frac{1}{X}$ with $\abs{\E[\hat{Q}] - 1/X} \leq \frac{\eps^{10}}{T}$ and $\Var[\hat{Q}] = O(\frac{1}{T^2}\log\frac{1}{\eps})$ for $N = O(\log^2 \frac{1}{\eps})$ for $\eps \in (0,1)$. 
\end{lemma}
\begin{proof}

Recall the Taylor series expansion for $f(x) = 1/x$ centered at $T$: 
\[
\frac{1}{x} = \sum_{i = 0}^\infty \frac{(-1)^{i} \cdot (x-T)^i}{T^{i+1}} =\frac{1}{T} - \frac{(x-T)^1}{T^2} + \frac{(x-T)^2}{T^3} - \ldots
\]
valid for $x \in (0, 2T).$  

Define $P_k(x)$ to be the associated degree $k$ Taylor polynomial. So, we have 
\[
P_k(x) = \sum_{i=0}^k \frac{(-1)^i \cdot (x - T)^i}{T^{i+1}}.
\]

Recall that when truncating a Taylor series at degree $k$ for some function $f(x)$, the error is 
\[
\abs{P_k(x) - f(x)} = \abs{\frac{f^{k+1}(\xi)}{(k+1)!} \cdot (x-T)^{k+1}}
\]
where $\xi$ is some fixed point between $T$ and $x$, and $f^{k+1}(\xi)$ denotes the value of the $(k+1)$-th derivative of function $f$ at $\xi$. 

For $f(x) = \frac{1}{x}$, we have that
\[
f^{k+1}(\xi) = \frac{(-1)^{k+1} \cdot (k+1)!}{\xi^{k+2}}.
\]
So, for $X \in [(1-\alpha) \cdot T, (1+\alpha) \cdot T]$, we have error 
\begin{align*}
\abs{P_k(X) - \frac{1}{X}}  &=  \frac{ (k+1)! }{\xi^{k+2} \cdot (k+1)!} \cdot \abs{(X -T)^{k+1}} 
= \frac{\abs{(X-T)^{k+1}}}{\xi^{k+2}} 
\\ &\leq \frac{(\alpha T)^{k+1}}{((1-\alpha) \cdot T)^{k+2}} = \frac{\alpha^{k+1}}{(1-\alpha)^{k+2} \cdot T}. \tag{since $X \in [(1-\alpha) \cdot T, (1+\alpha) \cdot T]$}
\end{align*}

By the lemma statement, we have access to $Y_1, \ldots, Y_N$ which are independent unbiased estimators of $X$ and therefore to $X-T$ after shifting by $T$. For each $Y_j$ for $j \in [N]$ we have 
\begin{align*}
\E[(Y_j - T)^2] & \leq \E[2(Y_j-X)^2 + 2(X-T)^2] = 2\Var[Y_j] + 2(X-T)^2 \tag{since $(A+B)^2 \leq 2A^2 + 2B^2$ by AM-GM} \\
& \leq \frac{T^2}{2} + 2(\alpha T)^2 \leq T^2. \tag{since $\Var[Y_j] \leq T^2/4$ and $\alpha < 1/2$}
\end{align*}
Let us take $\ell$ of the independent estimators $Y_j - T$ and multiply them. Call this estimator $Y^\ell$. Since we have that the estimators $Y_j - T$ for $j \in [N]$ are independent, we have that $\E[Y^\ell] = (X - T)^\ell$. Furthermore, we have that 
\[\E[(Y^{\ell})^2]  = \E[((Y_1 - T)(Y_2 - T)\ldots (Y_\ell - T))^2] \leq (T^2)^\ell.
\]

By simply rescaling, we get an estimator $Z_\ell$ such that \[\E[Z_\ell] = \frac{(-1)^\ell (X-T)^\ell}{T^{\ell+1}}, \E[Z_\ell^2] \leq \frac{(T^2)^\ell}{T^{2(\ell+1)}} = \frac{1}{T^2}.
\]

Let us set our estimator $\hat{Q} = \sum_{i=1}^k Z_i$. By using our upperbound on $\abs{P_k(X) - \frac{1}{X}}$ above, we have 
\[
\abs{\E[\hat{Q}] - \frac{1}{X}} = \abs{\E[Z_1 + \ldots + Z_k] - \frac{1}{X}} \leq \frac{\alpha^{k+1}}{(1-\alpha)^{k+2} \cdot T}.
\]

We also have 
\[
\Var[\hat{Q}] = \Var[Z_1 + \ldots + Z_k] = \Var[Z_1] + \ldots + \Var[Z_k] \leq \E(Z_1^2) + \ldots + \E(Z_k^2)
\leq \frac{k}{T^2}.
\]
The result follows upon setting $k = \Theta(\log \frac{1}{\eps}).$ 
\end{proof}
Therefore, we can conclude the following by multiplying the estimator for $\| \ff \|_p^p$ and $1/f_i^p$. 
\begin{lemma} \label{lem:sampProb}
Using $O(\log n \log^2(1/\eps))$ bits of space, for an index $i$ sampled by an $\ell_p$ sampler, we have an estimator $\hat{p}_i$ to $p_i = f_i^p / \| \ff \|_p^p$ with $\left|\E[1/\hat{p}_i] - 1/p_i\right| \leq \eps^{8} \cdot 1/p_i$ and $\E[1/\hat{p}_i^2] = O(1/p_i^2 \cdot  \log^2(1/\eps))$. 
\end{lemma}

\section{$F_p$ Estimation with Forgets}
\label{sec:forget_new}

\subsection{$F_p$ Estimation for $p \in (0, 2]$}
\label{sec:f_p2}
In this section, we will present algorithms for $F_p$ estimation ($0 < p \le 2$) in the presence of forgets. Let $\ff$ be the vector without the forget operations and $\bg$ be the actual frequency vector with forget operations. For the purpose of demonstration, we will first give an algorithm with $1/\eps^{3}$ dependence and show how to reduce this to $1/\eps^2$. At a high level, we run $k = \widetilde{O}\left({\frac{1}{\eps^2(1 - \alpha)}}\right)$ $\ell_p$ samplers from \Cref{sec:sampler}. Let $\mathcal{I}$ denote the set of sampled indices. If we are allowed to take a second pass on the data stream, then we can maintain an individual counter for each coordinate in $\mathcal{I}$, which means that we can get the extra value of $g_j$ for every $j \in \mathcal{I}$. On the other hand, we can also get an estimation of the reverse sampling probability $\hat{\frac{1}{p_i}}$ (\Cref{lem:sampProb}). One can argue that $\frac{1}{k} \cdot \sum_{j \in \mathcal{I}} g_j^p / \hat{p}_j$ will be a good estimator of the $F_p$ moment of $\bg$.

However, we are not allowed a second pass on the stream. To implement the algorithm in a single-pass stream, the observation is that if $g_i \ge (1 - \eps) f_i$, then we can just use $f_i$ as an estimation of $g_i$ as we are allowing a $(1 \pm \eps)$-approximation. On the other hand, when $g_i < (1 - \eps) f_i$, we have that there must be a forget operation after the frequency of coordinate $i$ in $\ff$ becomes $\eps f_i$. Let $\bz$ denote the corresponding rescaled vector of $\ff$ in the $\ell_p$ sampler. Recall that since $i$ is the sampled coordinate, we have $z_i = \Theta(1) \norm{\bz_{-i}}_2$. This implies that if we maintain the $\eps$-heavy hitters of $\bz$ during the stream and assign a counter to each of them. The corresponding counter can observe such a forget operation and then get the extra value of $g_i$. The above observation gives a one-pass streaming algorithm, but since for each $\ell_p$ sampler, we need an extra $\widetilde{O}\left(\frac{1}{\eps^2}\right)$ space for each sampler to keep track of the $
\eps$-heavy hitters, which result in a total $O(\eps^{-4})$ dependence. To get a better space, the crucial observation is that it is not necessary to maintain all the $\eps$-heavy hitters in all $\ell_p$ samplers. For example, if $g_i = 1/2 f_i$, then maintaining $\Theta(1)$-heavy hitter during the stream is sufficient.

Our algorithm is presented in~\Cref{fig:1}. At a high level, from each sub-routine in each sampler $L_i$ (recall that for each $\ell_p$, we have multiple groups of exponential random variables to make sure one of them passes the statistical testing), we uniformly sample $w_i \in (\eps, 1)$ and keep track of the $w_i$-heavy hitters of the rescaled frequency vector $\bz$. For sampled vector $j$, Our estimator for $g_j$ will be $X_j =  \mathbf{1} \left(\frac{g_j}{\hat{f}_j} \ge 1 - w_i \right) \cdot \hat{f}_j'$ where $\hat{f}_j$ and $\hat{f}_j'$ are two independent unbiased estimator of $f_j$. Intuitively, the expectation of $\mathbf{1} \left(\frac{g_j}{\hat{f}_j} \ge 1 - w_i \right)$ should be close to $\frac{g_j}{f_j}$ and then the expectation of $X_j$ is close to $g_j$. To prove the algorithm, we will need the following lemmas. 
\begin{figure}
    \begin{mdframed}
        \begin{algorithmic}
            Initlalize $k = \widetilde{O}\left(\frac{1}{\eps^2 (1-\alpha)}\right)$ $\ell_p$ sampler (\Cref{sec:sampler}) $L_1, L_2, \dots, L_k$ with parameter $w_i$ where $w_i$ is uniformly sampled in $(\eps, 1)$.
            
            For each subroutine in sampler $L_i$, let $\bz$ denote the frequency vector $\ff$ (without forget operations) rescaled by the exponential random variable. 
            
            \textbf{During the stream}: for each sub-routine in $L_i$, we use a BPTree (\Cref{lem:bpTHH}) to keep track of $(\frac{w_i}{10}, \ell_2)$-heavy hitter of $\bz$ and a continuous $F_2$ estimator (\Cref{lem:continious_l2}) to keep track of $F_2(\bz)$, whenever $\hat{F_2}(\bz)$ increases by a factor of $(1 + O(w_i))$:

            \begin{enumerate}
                \item Let $J$ be the set of heavy hitters output by the BPTree.
                \newline Discard all counters $C_i$ that are not associated with one coordinate in $J$.
                \newline For each $i \in J$, if there is no counter associated with $i$. Initialize a new counter $C_i = 0$ and use $C_i$ to keep track of the future updates to the coordinate $i$ in the stream. In particular, if there is $(i, +)$ operation, we increase $C_i$ by $1$, and if there is a forget operation, we set $C_i = 0$.

            \end{enumerate}
        \textbf{After the stream:} for each sampler $L_i$, let $j$ be the sampled coordinate with rescaled frequency vector $\bz$.   
            \begin{enumerate}
                \item Let $C_j$ be the counter associated with $j$ during the stream 
                \newline Let $\hat{f}_j$ be an estimation of $f_j$ and $\hat{f}_j'$ be another estimation of $f_j$.
                \newline Let $\hat{g}_j = C_j$ if $C_j$ observed a forget operation on $j$ during the stream after it was initialized or we set $\hat{g}_j = \hat{f}_j$ otherwise. 
                \newline Compute the value $X_j =  \mathbf{1} \left(\frac{\hat{g}_j}{\hat{f}_j} \ge 1 - w_i \right) \cdot \hat{f}_j'$ (\Cref{lem:delta_1}) and the reverse sampling probability $\hat{\frac{1}{p_j}}$ (\Cref{lem:sampProb}).
                \newline If $\hat{g_j}/\hat{f_j} \le \frac{1}{2}$, set $\hat{X_j^p} = \hat{g}_j^p$, otherwise compute $\hat{X_j^p}$ using the Taylor series (\Cref{lem:TSXP}).
                \newline 
            \end{enumerate}
            \textbf{Output:} the average of $\hat{X_j^p} \cdot \hat{\frac{1}{p_j}}$ for each $\ell_p$ sampler.
        \end{algorithmic}
    
    \end{mdframed}
    \caption{$F_p$ Estimation with Forget Operations}
    \label{fig:1}
\end{figure}
\begin{lemma}[Continuous $F_2$ estimator, \cite{BCINWW2017bptree}]
\label{lem:continious_l2}
    Let $0 < \eps < 1$. There is a streaming algorithm that outputs at each time $t$ a value $\hat{F}_2^{(t)}$ such that 
    \[
        \Pr \left(|\hat{F}_2^{(t)} - F_2^{(t)}|)\le \eps F_2^{(t)}, \ \text{for all $0 \le t \le m$} \right) > 1 - \delta \;.    
    \]
    The algorithm uses $O(\eps^{-2} \log n \log(1/\delta))$ bits of space.
\end{lemma}

\begin{lemma}
    \label{lem:tarlor_new}
    Using $O(w^{-2} \log n \log^3(1/\eps))$ bits of space, for an index $j$ sampled by an $\ell_p$ sampler, we have an estimator $\hat{f}_j$ to $f_j$ such that $\left|\E[1/\hat{f}_j] - 1/f_j\right| \le \eps^{8} \cdot 1/f_j$ and with probability at least $1 - \poly(\eps)$, $|f_j - \hat{f}_j| \le \frac{w}{100} \cdot f_j$.
\end{lemma}
\begin{proof}
    Let $\bz$ be the vector $\ff$ recaled by the exponential random variables $u_i^{1/p}$. We use a CountSketch with $O(w^{-2})$ buckets to estimate the frequency of the sampled index $z_j$ (this can be derandomized from the same procedure in~\cite{JW2021perfect}). By the property of CountSketch, we have $\E[\hat{z}_j] = z_j$. Note that we have that $|z_j|^2 \geq \Theta(1) \| \bz \|_2^2$ since $j$ is the sampled coordinate. Since we can use $\log(1/\eps)$-wise independence hash function for the random signs of CountSketch, from the standard concentration result, we have with probability $1 - \poly(\eps)$, $|\hat{z}_j - z_j| \le \frac{w}{100} z_j$. Multiplying by $u_j^{1/p}$ in both side we get the desired result.
    
    Moreover, by taking $\log^2 (1/\eps)$ independent copies and plug into the Taylor's estimator in~\Cref{lem:taylor} we can get a $\hat{f}_j$ such that $\left|\E[1/\hat{f}_j] - 1/f_j\right| \le \eps^{8} \cdot 1/f_j$. Since with probability at least $1 - \poly(\eps)$, each input here has a distance to $f_j$ within $\frac{w}{100}f_j$, condition on this, we have $|f_j - \hat{f}_j| \le \frac{w_i}{100} \cdot f_j$.
\end{proof}

\begin{lemma}
    \label{lem:delta_1}
    With probability at least $1 - \poly(\eps)$, we have that $\mathbf{1} \left(\frac{\hat{g}_j}{\hat{f}_j} \ge 1 - w_i \right) = \mathbf{1} \left(\frac{g_j}{\hat{f}_j} \ge 1 - w_i \right)$. 
    \begin{proof}     
        We first consider the case when $g_j \ge (1 - w_i/3) \cdot f_j$. In this case we have that $\mathbf{1} \left(\frac{g_j}{\hat{f}_j} \ge 1 - w_i \right) = 1$. Suppose that the counters for heavy hitters observe the last forget operations on $j$, then we can get the exact value of $g_j$. On the other hand, if this does not occur we will set $\hat{g}_j = \hat{f}_j$, which means that $\mathbf{1} \left(\frac{\hat{g}_j}{\hat{f}_j} \ge 1 - w_i \right)$ = 1.
        
        We next consider the other case $g_j < (1 - w_i/3) \cdot f_j$. This implies that there was a forget operation after the frequency of $f_j$ became $w_i f_j/ 3$ during the stream. Recall that in the corresponding rescaled frequency vector $\bz$ we have $z_j = \Theta(1) \norm{\bz_{-j}}_2$ and we check the $w_i/10$-heavy hitters of $\bz$ whenever the $F_2$ norm of $\bz$ increase by a $(1 + O(w_i))$-factor. This means that condition on the success of the continuous $F_2$ norm estimator and the BPTree we have that the coordinate $j$ will be identified as a heavy hitter before the last operation to $j$, and the counter will not be discarded in the reminder of the stream. This implies that $\hat{g}_j = g_j$.  
    \end{proof}
\end{lemma}

\begin{lemma}
    \label{lem:delta_2}
    With probability at least $1 - \poly(\eps)$, we have that $\E \left[\mathbf{1} \left(\frac{g_j}{\hat{f}_j} \ge 1 - w_i \right)\right] = \frac{g_j}{f_j}\left(1 \pm O(\eps)\right)$.
\end{lemma}
\begin{proof}
    We first consider the case when $g_j \ge (1 - c\eps) f_j$ for some small constant $c$. Note that since $w_i \in (\eps, 1)$ we have $\mathbf{1} \left(\frac{g_j}{\hat{f}_j} \ge 1 - w_i \right) = \frac{g_j}{f_j}\left(1 \pm O(\eps)\right)$. 
    
    We next consider the other case.
    Assume $g_j = (1 - v) f_j$, and condition on the event that $|f_j - \hat{f}_j| \le \frac{w_i}{100} \cdot f_j$. We have that if $w_i \le 0.9v$, then $g_j / \hat{f}_j \le (1 - v)\cdot(1 + \frac{w_i}{100}) < 1 - w_i$ and $g_j / f_j = (1 - v) < 1 - w_i$. This implies 
    \[
    \mathbf{1} \left(\frac{g_j}{\hat{f}_j} \ge 1 - w_i \right) = \mathbf{1} \left(\frac{g_j}{f_j} \ge 1 - w_i \right) = 0.
    \]
    Similarly, when $w_i \ge 1.1v$, we have that $g_j / \hat{f}_j \ge (1 - v)\cdot(1 - \frac{w_i}{100}) > 1 - w_i$ and $g_j / f_j = (1 - v) > 1 - w_i$. This implies 
    \[
    \mathbf{1} \left(\frac{g_j}{\hat{f}_j} \ge 1 - w_i \right) = \mathbf{1} \left(\frac{g_j}{f_j} \ge 1 - w_i \right) = 1.
    \]
    We next consider the case when $0.9v < w_i < 1.1v$. Let $\mathcal{D}$ denote this event, $p_{\mathcal{D}}$ denote the probability that $\mathcal{D}$ occurs, and $p_v$ denote the probability that $w_i \ge 1.1v$. Note that since $w_i$ is uniformly sampled in $(\eps, 1)$, we have
    \[
    \E\left[\mathbf{1} \left(\frac{g_j}{f_j} \ge 1 - w_i \right) \right] = \frac{g_j / f_j }{1 - \eps} \; .
    \]
    We then have 
    \[
    \E\left[\mathbf{1} \left(\frac{g_j}{f_j} \ge 1 - w_i \right) \ | \ \mathcal{D} \right] = \frac{(1-\eps)^{-1} g_j / f_j - p_v}{p_{\mathcal{D}}}
    \]
    from a simple probability computation. On the other hand, we have
    \begin{align*}
        \E\left[\mathbf{1} \left(\frac{g_j}{\hat{f}_j} \ge 1 - w_i \right) \ | \ \mathcal{D} \right] & = \E\left[\E\left[\mathbf{1} \left(\frac{g_j}{\hat{f}_j} \ge 1 - w_i \right) \ | \ \hat{f}_j, \mathcal{D}\right]\right] \\
        & = \E \left[ \frac{(1 - \eps)^{-1} g_j / \hat{f}_j - p_v}{p_{\mathcal{D}}}\right] = \frac{(1 - \eps)^{-1} g_j\cdot \E[1 / \hat{f}_j] - p_v}{p_{\mathcal{D}}}\\
        & = \frac{(1 - \eps)^{-1} (g_j / f_j)\cdot(1 \pm \eps^8) - p_v}{p_{\mathcal{D}}}
    \; .
    \end{align*}
    Here the last equation follows from~\Cref{lem:tarlor_new}. 

    Putting everything together, we have that $\E \left[\mathbf{1} \left(\frac{g_j}{\hat{f}_j} \ge 1 - w_i \right)\right] = \frac{g_j}{f_j}\left(1+ O(\eps)\right)$.
\end{proof}
\begin{lemma}
    \label{lem:X_j}
    With probability at least $1 - \poly(\eps)$, we have that $\E\left[X_j\right] = g_j \cdot \left(1 \pm O(\eps)\right)$ and $\E\left[X_j^2\right] = O(f_j \cdot g_j)$
\end{lemma}
\begin{proof}
    Recall that from~\Cref{lem:delta_1} we have that $X_j = \mathbf{1} \left(\frac{g_j}{\hat{f}_j} \ge 1 - w_i \right) \cdot \hat{f}_j'$ and $\hat{f}_j$ and $\hat{f}_j'$ are two independent random variables. Then from~\Cref{lem:delta_2} and~\Cref{lem:unbiased_est_of_fx} we have that with probability at least $1 - \poly(\eps)$,
    \[
    \E \left[ X_j \right] = \E \left[\mathbf{1} \left(\frac{g_j}{\hat{f}_j} \ge 1 - w_i \right)\right] \cdot \E\left[\hat{f}_j'\right] = \frac{g_j}{f_j} \cdot \left(1 \pm O(\eps)\right) \cdot f_j= g_j \cdot \left(1 \pm O(\eps)\right)
    \]
    and
    \[
    \E\left[X_j^2\right] = \E \left[\mathbf{1} \left(\frac{g_j}{\hat{f}_j} \ge 1 - w_i \right)^2\right] \cdot \E\left[\left(\hat{f}_j'\right)^2\right] = \frac{g_j}{f_j}\cdot\left(1 \pm O(\eps)\right) \cdot O(f_j^2 ) = O(f_j \cdot g_j) \;. \qedhere
    \]
\end{proof}
We are now ready to prove the following Theorem.
\begin{theorem}
    With high constant probability, the algorithm in~\Cref{fig:1} uses space $\widetilde{O}\left(\frac{1}{\eps^3(1 - \alpha)} \cdot \log n\right)$ for $0 < p < 2$ and $\widetilde{O}\left(\frac{1}{\eps^3(1 - \alpha)} \cdot \log^2 n\right)$ for $p = 2$, and outputs a value $\widetilde{F}$ with
\[
\abs{\tilde{F} - F_p(\bg)} \leq \eps F_p(\bg)
\]    
in the $\alpha$-RFDS model.
\end{theorem}

\begin{proof}
    As in~\Cref{fig:1}, we run $N = \widetilde{O}\left(\frac{1}{\eps^2 (1 - \alpha)}\right)$ $\ell_p$ sampler in~\Cref{thm:sampler}. At the end of the stream, we get $N$ sampled indices $i$ where the sampling probability is proportional to the value of $f_i^p$. For each sampled index $j$, when $\hat{g}_j /\hat{f}_j \le 1/2$, with probability at least $1 - \poly(\eps)$ we can get the exact value of $g_j$, otherwise from~\Cref{lem:X_j} we can get an estimator $X_j$ such that with probability at least $1 - \poly(\eps)$, $\E\left[X_j\right] = g_j \cdot \left(1 \pm O(\eps)\right)$ and $\E\left[X_j^2\right] = O(f_j \cdot g_j)$. Then, condition on this, from~\Cref{lem:TSXP} we can use it to compute an estimator $\hat{X_j^p}$ (we can run $O(\log(1/\eps))$ independent copies of $X_j$ and then plug into the Taylor's series) such that $\E\left[\hat{X_j^p}\right] = g_j^p \cdot \left(1 \pm O(\eps)\right)$ and $\E\left[\hat{X_j^{2p}}\right] = O(f_j^p \cdot g_j^p)$. Next, from~\Cref{lem:sampProb} we have that we can compute an unbiased estimator $\hat{p_j}$ such that $\left|\E[1/\hat{p}_j] - 1/p_j\right| \leq \eps^{8} \cdot 1/p_i$ and $\E[1/\hat{p}_j^2] = O(1/p_j^2 \cdot  \log^2(1/\eps))$, where $p_j = \frac{f_j^p}{\norm{\ff}_p^p}$. This implies that for the estimator $\hat{\frac{1}{p_j}} \cdot \hat{X_j^p}$ we have that 
    \[
\E\left(\hat{\frac{1}{p_j}} \cdot \hat{X_j^p}\right) = \sum_{i = 1}^n p_i \cdot \frac{1}{p_i}\left(1 + \eps^{10}\right) \cdot g_i^p\left( 1 \pm O(\eps)\right) = \left(1 \pm O(\eps)\right) \cdot  \sum_{i = 1}^{n} g_i^p = \left(1 \pm O(\eps)\right) \cdot F_p(\bg). 
\]
and
\[
\E\left(\hat{\frac{1}{p_j}} \cdot \hat{X_j^p}\right)^2 \le \widetilde{O}(1) \cdot \sum_{i = 1}^n p_i \cdot \frac{1}{p_i^2} \cdot (f_i g_i)^{p} \le \widetilde{O}(1) \cdot \sum_{i = 1}^n \frac{\norm{\ff}_p^{p}}{f_i^p} (f_i g_i)^{p} = \widetilde{O}(1) \cdot \norm{\ff}_p^p \cdot \sum_{i = 1}^n g_i^p \le \widetilde{O}\left(\frac{1}{1-\alpha} \right) \norm{\bg}_p^{2p}\;, 
\] 
where the $\widetilde{O}(\cdot)$ hides an $O(\log(1/\eps))$ factor and the last inequality follows from the assumption of the $\alpha$-RFDS model that $\norm{\bg}_p^p \ge (1 - \alpha) \norm{\ff}_p^p$. Taking a union bound over all of the $N$ samples, and by Chebyshev's inequality we have that after taking an average of all of the $N$ samples, with high constant probability, we have that 
\[
\abs{\tilde{F} - F_p(\bg)} \leq \eps F_p(\bg) \;.
\]
We next consider the space complexity of our algorithm. We first consider the case when $0 < p < 2$. There are $N = \widetilde{O}\left(\frac{1}{\eps^2 (1 - \alpha)}\right)$ $\ell_p$ samplers, for the $i$-th sampler $L_i$, since we need to keep track of the $w_i/10$-heavy hitters for each rescaled frequency vector $\bz$, the space usage will be $\widetilde{O}(w_i^{-2} \log n \cdot \poly \log(1/\eps))$. Hence, the total space usage will be 
\[
\widetilde{O}\left(\log n \cdot \poly \log(1/\eps) \cdot \sum_{i = 1}^N {w_i^{-2}} \right) \;.
\]
Recall that $w_i$ are uniformly sampled from $(\eps, 1)$. Then we have that $\E \left[\sum_{i = 1}^N {w_i^{-2}}\right] = O\left(1/\eps^3 \cdot \log \left(\frac{1}{\eps(1 - \alpha)}\right)\right)$, which implies the expectation of the space usage of our algorithm will be $\widetilde{O}\left(\frac{1}{\eps^3 (1 - \alpha)} \cdot \log n\right)$. For the case $p = 2$, note that the only difference here is for each $\ell_2$ sampler, we have $\widetilde{O}(\log n \log(1/\eps))$ sub-routines instead of $O(\log \log n \log(1/\eps))$ for $0 < p < 2$, which implies that with high probability, the expectation of the total space usage will be $\widetilde{O}\left(\frac{1}{\eps^3 (1 - \alpha)} \cdot \log^2 n\right)$. 
\end{proof}

\paragraph{Reducing to $1/\eps^2$ dependence.} Unfortunately, the above algorithm has a $1/\eps^3$ dependence. We next discuss how to achieve a tight $1/\eps^2$ dependence here. The key idea is that for each range of the value of $w_i$, we actually have more samples than we need. Specifically, consider a fixed level $[2^{\ell}\eps, 2^{\ell + 1}\eps]$ for some $\ell \in \{0, 1, \dots, \log(1/\eps)\}$. Since the total number of samples is $N = \widetilde{O}\left(\frac{1}{\eps^2 (1-\alpha)}\right)$, with probability at least $1 - \poly(\eps)$, we have the number of $w_i$ in this level is $N_\ell = O\left(2^{\ell}/\eps\cdot \frac{1}{1-\alpha} \right)$. What we do next is we randomly sample $N_\ell^{'} = O(4^{\ell}\cdot\frac{1}{1-\alpha})$ of them and assign each of these estimators with weight $\frac{N_\ell}{N_\ell^{'}}$. Clearly, after this step, our overall estimator is still unbiased. We next consider the variance of the new estimator. Note that for each of the $\ell_p$ sampler in this level, condition on its sampled index is $j$, we have $\E\left[X_j^2\right] = \E \left[\mathbf{1} \left(\frac{g_j}{\hat{f}_j} \ge 1 - w_i \right)^2\right] \cdot \E\left[\left(\hat{f}_j'\right)^2\right]\le \E\left[\left(\hat{f}_j'\right)^2\right] = O(f_j^2) = O(f_j \cdot g_j)$. The reason for the last equation is that we can, without loss of generality, assume $g_j \ge 1/3 \cdot f_j$ as otherwise we can get the exact value of $g_i$ directly. Then, similarly to the previous section, we can have $\E\left(\hat{\frac{1}{p_j}} \cdot \hat{X_j^p}\right)^2  \le \widetilde{O}\left(\frac{1}{1-\alpha} \right) \norm{\bg}_p^{2p}.$ From this we have that after the uniform sampling and rescaling, the variance of the sum of estimations in this level (without the averaging) will be at most
\[
\left(\frac{N_\ell}{N_\ell^{'}}\right)^2 \cdot N_\ell^{'} \cdot \widetilde{O}\left(\frac{1}{1-\alpha} \right) \norm{\bg}_p^{2p} = \widetilde{O}\left(\frac{1}{\eps^2} \cdot \frac{1}{\left(1-\alpha\right)^2} \right) \norm{\bg}_p^{2p} \;.
\]
Hence, after taking a sum over the $\log(1/\eps)$ levels and averaging by $N$, we have that the variance of the final estimator will be at most
\[
\frac{1}{N^2} \cdot \log(1/\eps) \cdot \widetilde{O}\left(\frac{1}{\eps^2} \cdot \frac{1}{\left(1-\alpha\right)^2} \right) \norm{\bg}_p^{2p} = O\left(\eps^2\right) \cdot \norm{\bg}_p^{2p} \;,
\]
which is the same order as before. We finally consider the space usage after this modification. For each level, we only need to maintain the $\ell_p$ samplers that have been uniformly sampled in this level, this implies the total space complexity will be
\[
\sum_{\ell = 0}^{\log(1/\eps)} N_\ell^{'} \cdot \frac{1}{2^{2\ell} \cdot  \eps^2} \cdot \widetilde{O} (\log^{t(p)} n \cdot \poly \log (1/\eps)) = \widetilde{O}\left(\frac{\log^{t(p)} n}{\eps^2 (1-\alpha)}\right) \;,
\]
where $t(p) = 1$ for $0 < p < 2$ and $t(p) = 2$ for $p = 2$. Putting everything together, we have the following theorem.

\begin{theorem}
\label{thm:fpopt1}
Let $0 < p \le 2$.
Let $\ff$ be the vector without the forget operations and $\bg$ be the actual frequency vector with forget operations.
There is an algorithm that uses space $\widetilde{O}\left(\frac{1}{\eps^2(1 - \alpha)} \cdot \log n\right)$ for $0 < p < 2$ and $\widetilde{O}\left(\frac{1}{\eps^2(1 - \alpha)} \cdot \log^2 n\right)$ for $p = 2$, and with high constant probability outputs a value $\widetilde{F}$ with
\[
\abs{\tilde{F} - F_p(\bg)} \leq \eps F_p(\bg)
\]    
in the $\alpha$-RFDS model.
\end{theorem}

\subsection{$F_p$ Estimation for $p > 2$}
We next consider the case $p > 2$. Here our strategy will be based on $\ell_2$ sampling. In particular, we consider the same algorithm as shown in~\Cref{fig:1} for $p = 2$ but using the Taylor series to estimate $\hat{X_j^p}$. Then, from a single estimator $\hat{\frac{1}{p_j}} \cdot \hat{X_j^p}$, it is easy to verify that the expectation is still almost unbiased and 
\[
\E\left(\hat{\frac{1}{p_j}} \cdot \hat{X_j^p}\right)^2 \le \widetilde{O}(1) \cdot \sum_{i = 1}^n p_i \cdot \frac{1}{p_i^2} \cdot (f_i g_i)^{p} \le \widetilde{O}(1) \cdot \sum_{i = 1}^n \frac{\norm{\ff}_2^{2}}{f_i^2} (f_i g_i)^{p} 
\]
To bound this, note that from H\"{o}lder's inequality we have $\norm{\ff}_2^2  \le n^{1-2/p} \norm{\ff}_p^2$ and we can assume $g_i \ge 1/3 f_i$ as otherwise we can get the exact value of $g_i$. Then we have 
\begin{align*}
\E\left(\hat{\frac{1}{p_j}} \cdot \hat{X_j^p}\right)^2 &\le \widetilde{O}(1)\norm{\ff}_2^2 \cdot \sum_{i = 1}^n f_i^{p - 2} g_i^p \le \widetilde{O}\left(1\right) \cdot 3^p \cdot \norm{\ff}_2^2 \cdot \norm{\bg}_{2p-2}^{2p - 2} \\
& \le \widetilde{O}\left(1\right) \cdot 3^p \cdot n^{1-2/p} \norm{\ff}_p^2 \cdot \norm{\bg}_{2p-2}^{2p-2} \le \widetilde{O}\left(\frac{3^p}{(1 - \alpha)^{2/p}}\right) \cdot n^{1 - 2/p} \cdot \norm{\bg}_p^2 \cdot \norm{\bg}_{2p-2}^{2p-2} \\
& \le \widetilde{O}\left(\frac{3^p}{(1 - \alpha)^{2/p}}\right) \cdot n^{1 - 2/p} \cdot \norm{\bg}_p^{2p} \;.
\end{align*}
This implies we only need to set the number of $\ell_2$ samplers to be $N = \widetilde{O}\left(\frac{1}{\eps^2 (1-\alpha)^{2/p}} \cdot n^{1-2/p}\right)$. Formally, we have the following theorem.
    
\begin{theorem}
\label{thm:fpopt}
Let $p > 2$ be a constant.
Let $\ff$ be the vector without the forget operations and $\bg$ be the actual frequency vector with forget operations.
There is an algorithm that uses $\widetilde{O}\left(\frac{1}{\eps^2(1 - \alpha)^{2/p}} \cdot n^{1 - 2/p}\cdot \log^2 n\right)$ bits of space, and with high constant probability outputs a value $\widetilde{F}$ with
\[
\abs{\tilde{F} - F_p(\bg)} \leq \eps F_p(\bg)
\]    
in the $\alpha$-RFDS model.
\end{theorem}

\subsection{$F_1$ Estimation}
\label{sec:F1}
In this section, we give a simple algorothm for $F_1$ estimation with forget operations. Note that $F_1$-estimation is straightforward without forget requests; we can simply keep a count of the number of insertions. To handle forget requests, alongside keeping a count of the number of insertions, we estimate the fraction of insertions that ultimately end up being deleted. We do this by sampling a uniform collection of $O(1/\eps^2)$ insertions via reservoir sampling and keeping track of the number of samples that are not deleted by forget requests. We note that the required counters in our algorithm naively require $O(\log m)$ bits where $m$ is the length of the stream. This turns out to be slightly suboptimal, and we address this minor issue with Morris counters to improve upon the logarithmic factors. 

The key procedure that we require is a version of reservoir sampling, slightly complicated by the fact that we only have room to approximately store the current stream length. 

\begin{lemma}
\label{lem:near_uniform_sampling}
There exists an algorithm which, given $\eps, \delta \in (0,1)$, an upper bound $M$ on the stream length, and an integer $k \geq 0$, samples $k$ insertion updates i.i.d. from the stream  using $O(k\log n + \log\log m + \log\frac{1}{\eps} + \log \log\frac{1}{\delta})$ space. If the stream has $m$ insertion updates then the probability that sample $j$ is update $i$ is 
$(1\pm \eps)\frac{1}{m}$ with probability at least $1-\delta$. 
\end{lemma}

\begin{proof}

We use a Morris counter to track the number of insertions.  To simply the analysis we use the Morris+ counter of Nelson and Yu~\cite{NY2022optimal}, which uses only $O(\log\log m + \log\frac{1}{\eps} + \log\log\frac{1}{\delta})$ space to estimate the count at any point to within $(1\pm \eps)$ multiplicative accuracy.  By replacing $\delta$ with $\delta/m$, we guarantee that the counter is accurate at all points in the stream simultaneously, with the same space requirements\footnote{This requires initially having an upper bound on the stream length.  However this is a very mild requirement, since this will later only contribute an additive $\log\log M$ factor.}. 

Now, we condition on the Morris+ counter having the stated guarantee.  That is, let $C(i)$ be the count after $i$ inserts into the stream.  We assume that we always have $(1-\eps) i\leq C(i) \leq (1+\eps)i$ for all $i$.  We show how to sample a single insert operation nearly uniformly.  Then we simply repeat the process $k$ times in parallel. 

We start by adding the first insertion update to our reservoir.  When we see the $i$-th insert operation, we swap it out for the current item in the reservoir with probability $1/C(i).$  If the $i$-th insert is added to the reservoir then the probability that it remains for the rest of the stream is
\[
\left(1 - \frac{1}{C(i+1)}\right)\left(1 - \frac{1}{C(i+2)}\right)\ldots \left(1 - \frac{1}{C(m)}\right).
\]

We will show that this product is close to $(1-\frac{1}{i+1})\ldots (1 - \frac{1}{m}).$  To do this, note that we have
\[
\frac{1 - \frac{1}{C(i+1)}}{1 - \frac{1}{i+1}}
\leq \frac{1 - \frac{1}{(1+\eps)(i+1)}}{1 - \frac{1}{i+1}}
= 1 + \frac{\eps}{(1+\eps) i}
\leq 1 + \frac{\eps}{i}.
\]

Similarly,
\[
\frac{1 - \frac{1}{C(i+1)}}{1 - \frac{1}{i+1}}
\geq \frac{1 - \frac{1}{(1-\eps)(i+1)}}{1 - \frac{1}{i+1}}
= 1 - \frac{\eps}{(1-\eps)i}
\geq 1 - \frac{2\eps}{i},
\]
for $\eps \leq \frac12$ \footnote{Note that requiring $\eps \leq 1/2$ still makes the algorithm applicable to the entire range of $\eps$. Given some $\eps > 1/2$, we can simply run the algorithm with $\eps / 2$, satisfying the error guarantee and only increasing the space by a constant factor.}.

Now let 
\[
P_m = \prod_{j=1}^m \left(1 + \frac{\eps}{j}\right).
\]
Then
\[
\ln (P_m) = \sum_{j=1}^m \ln\left(1 + \frac{\eps}{j}\right).
\]
By the bound $\frac{1}{2}x \leq \ln(1+x) \leq x$ valid for $0\leq x \leq 2$ we have 
\begin{align*}
\frac{1}{2}\eps\ln m &\leq \sum_{j=1}^m \frac12 \frac{\eps}{j} \tag{by bounds on the Harmonic number}\\&\leq \ln (P_m) \leq \sum_{j=1}^m \frac{\eps}{j} \tag{by taking $x = \frac{\eps}{j}$}
\\& \leq \eps (\ln(m) + 1)
\leq 2\eps\ln m. \tag{by bounds on the Harmonic number}
\end{align*}
So, we have that 
\begin{equation}
    \frac{1}{2}\eps \ln m \leq \ln (P_m) \leq 2 \eps \ln m.
\end{equation}

Similarly set $Q_m = \prod_{j=1}^m \left(1 - \frac{2\eps}{j}\right),$ so that
\[
\ln( Q_m) 
= \sum_{j=1}^m \ln \left(1 - \frac{2\eps}{j}\right).
\]
Note that we have the bound $-2x \leq \ln (1 - x) \leq -x$ for $0\leq x \leq \frac{1}{2}.$
So we have
\begin{align*}
-4\eps\ln m & \leq \sum_{j=1}^m -\frac{4\eps}{j}
\\ &\leq  \ln( Q_m) \leq -2\eps (\log m + 1) \leq -4\eps\log m.
\end{align*}

It follows from these bounds that
\begin{align*}
\left(1 - \frac{1}{C(i+1)}\right)\left(1 - \frac{1}{C(i+2)}\right) \ldots \left(1 - \frac{1}{C(m)}\right)
&\leq \frac{P_{m-1}}{P_{i-1}} (1 + \frac{1}{i+1})\ldots (1 + \frac{1}{m})\\
&= \frac{P_{m-1}}{P_{i-1}} \frac{i}{m}\\
&\leq \exp(2\eps\log m) \frac{i}{m}.
\end{align*}
Similarly 
\begin{align*}
\left(1 - \frac{1}{C(i+1)}\right)\left(1 - \frac{1}{C(i+2)}\right)\left(1 - \frac{1}{C(m)}\right)
&\geq \frac{Q_{m-1}}{Q_{i-1}} (1 + \frac{1}{i+1})\ldots (1 + \frac{1}{m})\\
&= \frac{Q_{m-1}}{Q_{i-1}} \frac{i}{m}\\
&\geq \exp(-4\eps\log m) \frac{i}{m}.
\end{align*}

It follows from this calculation that insert operation $i$ is sampled with probability at least 
\[
\frac{1}{C(i)}\exp(2\eps \log m)\frac{i}{m} \geq \frac{1}{1+\eps}\exp(2\eps \log m)\frac{1}{m}
\geq (1+\eps) \exp(2\eps \log m)\frac{1}{m},
\]
and probability at most
\[
\frac{1}{C(i)}\exp(-4\eps \log m)\frac{i}{m}
\leq \frac{1}{1-\eps}\frac{1}{i}\exp(-4\eps \log m)\frac{i}{m}
\leq (1+2\eps) \exp(-4\eps \log m) \frac{1}{m}.
\]
Finally, replace $\eps$ with $\eps/\log m$ to obtain the stated guarantee.
\end{proof}

\begin{theorem}
\label{thm:F_1}
There is an algorithm that, given $\eps \in (0,\frac12)$, $\alpha, \delta \in (0,1)$, obtains a $(1\pm \eps)$ approximation to $F_1$ in the $\alpha$-RFDS model using space
$O(\frac{1}{1-\alpha}\frac{1}{\eps^2}\log n\log\frac{1}{\delta}).$ 
\end{theorem}

\begin{proof}
The first step is to obtain a $1\pm (1-\alpha)\eps$ approximation to the total number of insert operations.  This is accomplished with a Morris counter using $O(\log\frac{1}{1-\alpha} + \log \frac{1}{\eps} + \log \log \frac{1}{\delta} + \log \log m)$ space. Recall that we have $m = \poly{n}$.

Next we will estimate the number of insert operations that are later deleted, which we do using our near-uniform sampling procedure from \Cref{lem:near_uniform_sampling} above. For each update that we sample, we additionally keep an extra bit that denotes whether that update is later forgotten (when a forget request comes, we simply check which of insert samples it affects). If an operation is deleted from our reservoir we delete its identity as well as the corresponding bit. 

Suppose that there are $m$ insert operations total and that $d = \beta m$ of them are ultimately not deleted. The probability that a single sample of our procedure chooses one of these $d$ insertions is in $[(1 -\eps)\beta, (1 +\eps)\beta]$ by \Cref{lem:near_uniform_sampling}. Let us suppose that we take $k$ samples using \Cref{lem:near_uniform_sampling}. Define $X_i$ for $i \in [k]$ to be an indicator random variable that equals $1$ if the $i$-th sample is one of the $d$ insert operations and $0$ otherwise. Take $X = \sum_{i=1}^k X_i$. We have $\E[X_i] \in (1 \pm \eps) \beta$ and $\E[X] \in k \cdot (1 \pm \eps) \beta$. Using that $\eps < 1$ along with a Chernoff bound we get 
\begin{align*}
\Pr(\abs{X - \beta k} \geq 3\eps \beta k)
&\leq \Pr(|X - \E[X]| \geq \eps \E[X])\\
&\leq 2 \cdot \exp\left(-\frac{\eps^2 \E[X]}{3} \right) \\
&\leq 2 \cdot \exp\left(-\frac{\eps^2 k \cdot (1- \eps) \beta}{3} \right)\\
&\leq \delta
\end{align*}
Note that $\beta \geq 1-\alpha$ by the $\alpha$-RFDS property.  So for $k = O(\frac{1}{(1-\alpha)\eps^2} \log\frac{1}{\delta})$ we have $\frac{m}{k}X \in \beta m \pm 3\eps \beta m$ with probability at least $1-\delta$. 
 
The $F_1$ moment of the stream after forget operations is $\beta m$, so we obtain a $(1\pm 3\eps)$-approximation to $F_1.$ The result follows after adjusting $\eps$ by a constant.
\end{proof}

\section{Extensions} \label{sec:extensions}

\subsection{General Operations} \label{subsec:GO}
In this section, we will first present and analyze our heavy-hitters data structure. Then, we will give our algorithm for $F_p$ estimation for $p > 0$ based on this heavy-hitters data structure. 
In particular, here we are allowing a group of more general functions including the forget operation. For a class of functions $\mathcal{G}$ which we will describe, at each time $t$ in the stream, we allow an update of the normal addition $(i, +)$ or of the type $(i, g_t \in \mathcal{G})$ which performs $f_i = g_t(f_i)$. For ease of presentation, we first work with $\mathcal{G}$ as the class of contracting functions, or functions with $g : \mathbb{N} \to \mathbb{N}$ with the following conditions: 
\begin{enumerate}
    \item $|g(x) - g(y)| \le |x - y|$
    \item $g(0) = 0$.
\end{enumerate}
We note that the actual class of functions $\mathcal{G}$ that our algorithm can accommodate is more general and we will formally define it below. 
To demonstrate our techniques, we first consider the case of $\ell_2$-heavy hitters. We will then show that it can be extended to the general $\ell_p$ case for $p \geq 2$. 

\subsubsection{$\ell_2$ Heavy-Hitters.} \label{subsub:HH}
Let $\widetilde{\ff}$ be the vector without the operations in $\mathcal{G}$ and $\ff$ be the actual frequency vector with the operations in $\mathcal{G}$. To find (and estimate) all the $\ell_2$ heavy hitters of $\ff$ it suffices to estimate all entries $f_i$ of $\ff$ to within $\eps \norm{\ff}_2$ additive error.  We give an algorithm for that here.

Let $\widetilde{\ff}^{(k)}$ be the underlying frequency vector after $k$ increments (not including general operations). Let $\tilde{}\ff^{(k)}$ be the corresponding underlying frequency vector if all general operations (i.e. those that are not increments) are removed from the stream. Let $m$ be the length of the stream. Notationally, we set $\ff := \ff^{(m)}$. 
By the $\alpha$-RFDS assumption, we have  $\norm{\ff}_2^2 \leq \norm{\tilde{\ff}}_2^2 \leq \frac{1}{1-\alpha}\norm{\ff}_2^2.$ Since our goal is $\eps \| \ff \|_2$ additive error, $\eps \sqrt{1-\alpha} \| \tilde{\ff}\|_2$ additive error suffices. Below we set $\eps' = \eps \sqrt{1-\alpha}$.

We now outline the data structures which we use in our heavy hitters algorithm. They are as follows: 
\begin{enumerate}
    \item An $\ell_2$ estimation sketch giving a $(1 \pm \eps)$  approximation to $\norm{\tilde{\ff}^{(k)}}_2$ accurate for all $k$ (\Cref{lem:continious_l2}). 

    \item A sketch that allows us to obtain all $\left(\frac{\eps'}{10}, \ell_2\right)$-heavy hitters of $\tilde{\ff}^{(k)}$ at the time point $k$. The sketch can be taken to be a BPTree at~\Cref{sec:bptree}.

    \item Counters $C_1, \ldots, C_n$.  Each counter can be either active or inactive.  We only store the active counters. Initially, they are all inactive. 
\end{enumerate}

Our algorithm handles two types of updates.  To process $(i,g)$ (meaning apply $g$ on coordinate $i$) we simply set the $C_i$ counter to $g(C_i)$ (and don't change its active status).  To process the $k$-th increment $(i, +)$ we do the following:
\begin{enumerate}
    \item Update the $\ell_2$ estimation sketch and the heavy hitter sketch.
    \item Let $M_k$ be an estimation of $\norm{\tilde{\ff}^{(k)}}_2$. Whenever $M_k$ increases by a $(1 + O(\eps))$-factor, query the heavy hitter sketch to get a set of heavy hitters $J_k$.
    \item For all $j \in J_k$, set $C_j$ to active.  Set all others to be inactive and zero them out.
    \item If $C_i$ is active, increment $C_i$.
\end{enumerate}

After the stream, to respond to an entry query in position $j$, we simply report the value of $C_j.$

\begin{lemma}
At the end of the stream, $|f_j - C_j| \leq \eps'\norm{\tilde{\ff}}_2.$
\end{lemma}
\begin{proof}
We first consider the indices $j$ such that we have $C_j$ is inactive at the end of the stream. From the way our algorithm processes increments, we have that 
\[
|f_j - 0| = f_j \leq \eps' \| \tilde{\ff}\|_2. 
\]
We next consider indices $j$ such that $C_j$ is active at the end of the stream. Suppose that the last time $C_j$ has been set to active is $T$. We will show that for every time $t = T + i$, we have $|f_j^{(t)} - C_j^{(t)}| \leq \eps'\norm{\tilde{\ff}}_2$.

We prove this by induction. For $i = 0$, since $C_j$ was just to be set active, then from the rules of contracting functions we have 
\[
|C_j^{(T)} - f_j^{(T)}| \le f_j^{(T)} \le f{'}_j^{(T)} \le \frac{1}{4}\eps' \norm{\tilde{\ff}^{(T)}}_2 \le \frac{1}{4}\eps' \norm{\tilde{\ff}}_2\]
which gives us $|f_j^{(T)} - C_j^{(T)}| \le \eps' \norm{\tilde{\ff}}_2$.

Suppose that for $i \le k - 1$ the inductive hypothesis is true. We next consider the case $i = k$. If at this time the operation is $(j, +)$, then $|f_j^{(T + k)} - C_j^{(T + k)}|$ will not change since both of the quantities will increase by $1$. On the other hand, it at this time the operation in the stream is $(j,g)$ for some function $g$, then we will have 
 \begin{equation} \label{eq:inducGO}
 |f_j^{(T + k)} - C_j^{(T + k)}| = \left|g(f_j^{(T + k - 1)}) - g(C_j^{(T + k - 1)})\right| \le |f_j^{(T + k - 1)} - C_j^{(T + k - 1)}| \le \eps' \norm{\tilde{\ff}}_2,
 \end{equation}
 which is what we need. 
\end{proof}

\begin{lemma}
The algorithm above uses $O\left(\frac{1}{(\eps')^2} \log n \log(1/\delta)\right)$ space. 
\end{lemma}
\begin{proof}
From~\Cref{lem:continious_l2}, we can ensure that our $\ell_2$ estimator is correct on all iterations with failure probability $\delta$ using $O(\frac{1}{\eps'^{2}}\log n \log(1/\delta))$ space.  
The space of the BPT will be $\widetilde{O}(\frac{1}{\eps'^2}) \log n \log(1/\delta)$ from~\Cref{lem:bpTHH}.

Recall that the $J_k$ returned by the BPTree has size $|J_k| = O(1/\eps'^{2})$. This means that each time during the stream, the number of active counters will be at most $O(1/\eps'^{2})$. Storing the identities of all the counters takes $O(\log n/(\eps')^2)$ space, and storing each of their values takes $O(\log n)$ space.  So the counters can be implemented with $O(\log n/(\eps')^2)$ space.
\end{proof}

\begin{theorem}
\label{thm:l2_hh}
There is an algorithm that estimates all coordinates of $\ff$ to within $\eps\norm{\ff}_2$ additive error using $O\left(\frac{1}{1-\alpha}\frac{1}{\eps^2} \log n \log(1/\delta)\right)$ space in the $\alpha$-RFDS model with general arbitrary contraction operations.
\end{theorem}
\begin{proof}
By setting $\eps' = \eps\sqrt{1-\alpha}$ as described above, our algorithm gives a way to estimate all coordinates of $v$ to within additive error $\eps\norm{\ff}_2$ additive error, using $O\left(\frac{1}{1-\alpha}\frac{1}{\eps^2}\log n \log(1/\delta)\right)$ space.
\end{proof}

We remark that in the above proof we assume that each function $g \in \mathcal{G}$ satisfies $|g(x) - g(y)| \le |x - y|$. This condition can be more extended. In fact, if an arbitrary sequence $g_1, g_2, \dots, g_k \in \mathcal{G}$ satisfies \Cref{eq:inducGO}, then our algorithm will be support for $\mathcal{G}$. On the other hand, if we can find a sequence in $\mathcal{G}$ that breaks this condition, then one can prove a lower bound via a reduction to the indexing in communication complexity.

\begin{corollary}
Suppose that $\mathcal{G}$ is a function class such that for every positive integers $x, k \le \poly(n)$ we can not select a function sequence $g_1, g_2, \dots, g_k \in \mathcal{G}$ such that 
\[
|g_k \circ g_{k - 1} \circ \dots \circ g_1(x) - g_k \circ g_{k - 1} \circ \dots \circ g_1(0)| > c x \
\]
for some absolute constant $c$. Then, there is an algorithm that estimates all coordinates of $\ff$ to within $\eps\norm{\ff}_2$ additive error using $O\left(\frac{1}{1-\alpha}\frac{1}{\eps^2} \log n \log(1/\delta)\right)$ space in the $\alpha$-RFDS model with general arbitrary contraction operations.

\end{corollary}
\subsubsection{$\ell_p$ Heavy-Hitters}
Next we show how to solve the $\ell_p$ heavy-hitters problem for $p \ge 2$ using essentially the same procedure as for $\ell_2$. 

\begin{theorem}
\label{thm:ell_p_heavy_hitters}
Let $p > 2$. In the $\alpha$-RFDS model with general operations from $\mathcal{G}$, there is a streaming algorithm to find all indices $i$ with $f_i \geq \eps \norm{\ff}_p$ with high probability using $\widetilde{O}(\frac{1}{\eps^2} \frac{1}{(1-\alpha)^{2/p}} n^{1 - 2/p})$
space with high probability.  Moreover this algorithm gives an additive $\eps\norm{\ff}_p$ approximation to the values of all heavy hitters.
\end{theorem}
\begin{proof}
As in the previous section, let $\tilde{\ff}$ be the final frequency vector without any general operations (i.e. only increments), and let $\ff$ be the actual underlying frequency vector.

So if $i$ is an ($\eps$, $\ell_p$)-heavy-hitter index, then
\[
f_i^p \geq \eps^p\norm{\ff}_p^p \geq \eps^p (1-\alpha)\norm{\tilde{\ff}}_p^p.
\]
Therefore
\[
f_i^2 \geq \eps^{2}(1-\alpha)^{2/p}\norm{\tilde{\ff}}_p^2
\geq \eps^2 (1-\alpha)^{2/p} n^{-1 + 2/p}\norm{\tilde{\ff}}_2^2.
\]

Set $\eps' = \eps (1-\alpha)^{1/p} n^{-1/2 + 1/p}$ so that when $i$ is an $\ell_p$-heavy-hitter, then $f_i \geq \eps' \norm{\tilde{\ff}}_2.$ The previous section gives an algorithm to estimate all coordinates of $\ff$ to within $\eps'\norm{\tilde{\ff}}_2$ additive error.  The above calculation shows that $\eps'\norm{\tilde{\ff}}_2 \leq \eps\norm{\ff}_p$, so we obtain an $\eps\norm{\ff}_p$ additive approximation to all heavy-hitter coordinates.
\end{proof}

We finally consider $0 < p < 2$. Note that in this case, the error of the $\ell_2$ heavy hitter algorithm for each coordinate will be $\eps' \norm{\tilde{\ff'}}_2 \le \eps' \norm{\tilde{\ff'}}_p \le \eps' \cdot \frac{1}{(1 - \alpha)^{1/p}}\norm{\ff}_p$. Hence, it is suffice to set $\eps' = \eps \cdot (1 - \alpha)^{1/p}$. We have the following theorem.

\begin{theorem}
\label{thm:ell_p_heavy_hitters2}
Let $0 < p < 2$. In the $\alpha$-RFDS model with general operations from $\mathcal{G}$, there is a streaming algorithm to find all indices $i$ with $f_i \geq \eps \norm{\ff}_p$ with high probability using $\widetilde{O}(\frac{1}{\eps^2} \frac{1}{(1-\alpha)^{2/p}} \cdot \log n)$ bits 
space.  Moreover this algorithm gives an additive $\eps\norm{\ff}_p$ approximation to the values of all heavy hitters.
\end{theorem}

\subsubsection{$F_p$ Estimation for $p > 0$}

In this section, we give the algorithm for $F_p$ estimation when stream updates can be general operations from $\mathcal{G}$. Our algorithm is inspired by classical level set arguments in the literature, which requires the following heavy hitter data structure as a sub-routine.

\begin{definition}
\label{def:HH}   
Given parameters $p, \theta, \eps$, a heavy hitter data structure $\mathcal{D}$ receives a stream of updates to the frequency vector $\ff$, and provides an estimate $\widetilde{\ff}$ to $\ff$ where, with probability at least $1 - 1/\poly(n)$, we have for all $i \in [n]$:
\begin{enumerate}
    \item $||f'_i|{^p} - |f_i|^p| \le \eps \cdot |f_i|^p$ if $|f_i|^p \ge \theta \|\ff\|_p^p$.
    \item $|f_i'|^p < \theta (1 +\eps) \cdot\|\ff\|_p^p$ if $|f_i|^p < \theta \|\ff\|_p^p$ 
\end{enumerate}

\end{definition}
Our \Cref{alg:levelSet} essentially follows from  Algorithm~3 in~\cite{LWY2021exponentially} but has a better $\eps$ dependence. 
Take $M$ to be an upper bound on the stream length. At a high level, we divide our vector's entries into level sets.  Namely, let $S_j$ denote the set of coordinates $f_i$ where $f_i\in [\frac{M}{2^{j + 1}}, \frac{M}{2^j}]$ at the end of the stream. We then estimate the $s_j = \sum_{i \in S_j} |f_i|^p$ individually. Then the final estimation becomes
\[
s = \sum_i |f_i|^p = \sum_j \sum_{i \in s_j} |f_i|^p = \sum_j s_j \;.
\]

For the case where $s_i \le \eps F_p / \log M$, we can treat this part as $0$ as it contributes error at most $\eps F_p / \log M$ and there are at most $\log M$ such $s_i$. For the other case where $s_i \ge \eps F_p / \log M$, we consider the sub-sampling scheme where we sub-sample the coordinates of the underlying vector with probability $p_i = 2^{-i}$ for $i = 1, 2,\cdots, \log M$. Consider the sub-sampled stream where there are $O(1/\eps^2)$ survivors which are the coordinates in $S_j$. From the condition $s_i \ge \eps F_p / \log M$ we get that these coordinates are $O(\eps^3/\log^2 M)$-heavy hitters in the corresponding sub-stream with high probability. Hence, with an $O(\eps^3/\log^2 M)$-heavy hitter data structure we can detect all of them with high probability. On the other hand, since we have $O(1/\eps^2)$ samples from $S_i$ and the coordinates in $S_i$ are within a factor of $2$, by Chebyshev's inequality we have that we can get a $(1 \pm \eps)$-approximation of each $s_i$, which yields a $(1 \pm \eps)$-approximation to $F_p$. For more details of the proof, we refer the readers to Section~5.5 in \cite{LWY2021exponentially}.

To obtain a better $\eps$ dependence, an observation we make is if $s_i$ is small, we will not need to compute a $(1 + \eps)$-approximation of $s_i$. For example, if $s_i = \Theta(\eps F_p)$, then a $O(1)$-approximation is sufficient.  In particular, in our algorithm we first look at the sub-stream level where there are $O(1)$-survivors in $S_i$, so we can get a constant approximation of each $s_i$ (or we will know $s_i < \eps F_p / \log M$). Then suppose that $s_i \approx \eps_i F_p/\log M$. In this case, we only need a $1 + \eps/\eps_i$-approximation of $s_i$, which means we can look at the sub-stream where they are $\Theta(\eps_i^2 / \eps^2)$ survivors in $S_i$, which means that a $\eps^2 / \log^2 M$ Heavy Hitter structure is sufficient.  

\begin{algorithm} [ht]
\caption{Estimating $F_p$ $(p > 0)$}
\label{alg:levelSet}
\begin{algorithmic}[1]
    \REQUIRE $(i)$ A sub-sampling scheme $H$ such that the $i$-th level has sub-sampling probability $p_i = 2^{-i}$ to each coordinates of the underlying vector; $(ii)$: $L + 1$ HeavyHitter structures $\mathcal{D}_0, \ldots ,\mathcal{D}_{L}$ with parameter $(p, \theta = \eps^2 / L^2, \eps)$ where $\mathcal{D}_i$ corresponds to the $i$-th sub-stream $L = O(\log M)$; $(iii)$: $M$ is an upper bound on the stream length .
    \STATE $j_0 \gets K \log(\eps^{-2} L^2)$ where $K$ is a large constant.
    \STATE $\zeta \gets$ uniform random variable in $[1/2, 1]$.
    \FOR{$j = 0, 1, \cdots, L$}
        \STATE $\Lambda_j \gets O(\eps^2 / L^2)$-heavy hitters in $\mathcal{D}_j$ 
    \ENDFOR
    
    \FOR{$j = 0, \ldots, j_0$}
        \STATE Let $\lambda_1^{(j)},\dots,\lambda_s^{(j)}$ be the elements in $\Lambda_0$ contained in $[(1+\eps)\zeta\frac{M}{2^j}, (2-\eps)\zeta\frac{M}{2^j}]$\;
        \STATE $\widetilde{S_j} \gets 
        |\lambda_1^{(j)}|^p+\cdots+|\lambda_s^{(j)}|^p$\;
 
    \ENDFOR
    \FOR{$j = j_0 + 1, \ldots, M$}
        \STATE Find the largest $\ell$ where $\Lambda_\ell$ contains $\Theta(1)$ elements $\lambda^{(j)}_1,\dots,\lambda^{(j)}_s$ in $[(1+\eps)\zeta\frac{M}{2^j}, (2-\eps)\zeta\frac{M}{2^j}]$
        \IF{such $\ell$ exists}
        \STATE $\widetilde{S_j} \gets (|\lambda_1^{(j)}|^p+\cdots+|\lambda_s^{(j)}|^p) \cdot 2^\ell$
        \ELSE
        \STATE $\widetilde{S_j} \gets 0$
        \ENDIF
    \ENDFOR
    \STATE $\widetilde{S}\gets \sum_j \widetilde{S}_j$\; \hfill $\triangleright${\color{purple} We first get a $O(1)$ approximation $\widetilde{S}$ to $F_p$}
    \FOR{$j = j_0 + 1, \ldots, M$}
        \IF{$\widetilde{S_j} \ne 0$}
        \STATE $\eps_j \gets \widetilde{S_j} \log M / \widetilde{S} $
        \STATE Find the largest $\ell$ where $\Lambda_\ell$ contains $\Theta\left(\frac{\eps_j^2  }{\eps^2}\right)$ elements  $\lambda^{(j)}_1,\dots,\lambda^{(j)}_s$ in $[(1+\eps)\zeta\frac{M}{2^j}, (2-\eps)\zeta\frac{M}{2^j}]$  
        \STATE $\widetilde{S_j} \gets (|\lambda_1^{(j)}|^p+\cdots+|\lambda_s^{(j)}|^p) \cdot 2^\ell$
        \ENDIF
    \ENDFOR
    \STATE \textbf{return} $\widetilde{S}\gets \sum_j \widetilde{S}_j$\; 
\end{algorithmic}
\end{algorithm}

\begin{lemma}[Essentially Section~5.5 in~\cite{LWY2021exponentially}]
    Given the heavy hitter data structure in \Cref{def:HH}, with high constant probability \Cref{alg:levelSet} returns a $(1 \pm \eps)$-approximation of $F_p$.  
\end{lemma}
We next show that our heavy hitter algorithm actually gives the implementation of the heavy hitter data structure we need.
Consider the heavy hitter data structure in \Cref{subsec:GO} with parameters $\eps' = \theta^{1/p} \eps = \eps^{2/p + 1} / L^{2/p}$. Then, consider the coordinate $i$ of the frequency vector $\ff$ where $|f_i|^p \ge \theta \|\ff\|_p^p$. This means we have $|f_i| \ge \theta^{1/p} \|\ff\|_p$. Then, from the definition of $\eps'$ we can get that the estimation $f_i'$ by the data structure is a $(1 \pm \eps)$-approximation of $f_i$. This means that $||f'_i|{^p} - |f_i|^p| \le O(\eps) \cdot |f_i|^p$. On the other hand when $|f_i|^p < \theta \|\ff\|_p^p$, similar from the guarantee of the data structure we have $|f_i'|^p < \theta (1 + \eps) \|\ff\|_p^p$. Combining these two things, we can get that the requirements of \Cref{def:HH} are satisfied, which yields the following theorem.

\begin{theorem}
\label{thm:F_p}
    There is an algorithm to obtain a $(1\pm \eps)$ approximation to $F_p$ for $p > 0$ in the $\alpha$-RFDS model with general operations from $\mathcal{G}$ using 
$\widetilde{O}\left(\frac{1}{(1-\alpha)^{2/p}}\frac{1}{\eps^{2 + 4 / p}} n^{1 - 2/p} \right)$ bits of space for $p > 2$ and $\widetilde{O}\left(\frac{1}{(1-\alpha)^{2/p}}\frac{1}{\eps^{2 + 4 / p}} \cdot \poly \log{n}\right)$ bits of space for $0 < p \le 2$.
\end{theorem}

\subsection{Prefix/Suffix-Deletion Models}
In the prefix-deletion model we allow for standard stream updates along with prefix deletion requests of the form $\prefixdelete(i, x)$ which deletes all occurrences of $x$ at time $i$ and prior.  We require that $\prefixdelete(i, x)$ is received at time $i$ or later.

In the suffix-deletion model we allow for standard stream updates along with suffix deletion requests of the form $\suffixdelete(i, x)$ which deletes all occurrences of $x$ at times in $[i,T]$ where $T$ is that at which the request is received. We will assume that no additional updates are received after a suffix deletion request.  We show how to modify our algorithms for $F_p$ estimation to this model.

As in the RFDS model, we will assume that the prefix and suffix deletions cause the statistic of interest to drop by at most a $(1-\alpha)$ factor.  We refer to this as the $\alpha$-PSD model, short for ``prefix-suffix deletion".

For $F_1$ estimation, the algorithm is essentially the same as in the RFDS model, and we obtain the same guarantee.
\begin{theorem}
\label{thm:F1_psd_model}
    There is an algorithm to obtain a $(1 \pm  \eps)$ approximation to $F_1$ in the $\alpha$-PSD model using
$O\left(\frac{1}{1-\alpha}\frac{1}{\eps^2}\log n\log\frac{1}{\delta} + \log\log m\right)$ bits of space. 
\end{theorem}

\begin{proof}
We use the same reservoir-sampling based algorithm as for $F_1$ estimation in the forget model to sample $\frac{1}{1-\alpha}\frac{1}{\eps^2}$ stream updates.  We simply need a way to decide when a given stream update is later erased by a request.  For this, we simply store the times at which we sample the stream updates.  Essentially the same algorithm applies in the suffix-deletion model.
\end{proof}

To adapt our $F_p$ moment for $p \geq 2$ estimation algorithms to allow for prefix and suffix deletions, it suffices to adapt our $\ell_2$ heavy-hitters algorithm, since we apply this as a black box to obtain $F_p$ moment estimators.

\begin{theorem}
\label{thm:l2_hh_PSD_model}
There is an algorithm that estimates all coordinates of $\ff$ to within $\eps\norm{\ff}_2$ additive error using $O\left(\frac{1}{1-\alpha}\frac{1}{\eps^2}(\log m \log\frac{m}{\delta} + \log n)\right)$ space in the $(\alpha, F_2)$-PSD model.
\end{theorem}
\begin{proof}
We borrow the notation from \Cref{subsub:HH}. Recall that our $\ell_2$-heavy-hitters algorithm in the forgot model maintains an estimate $M^{(k)}$ of the current $\ell_2$ value (ignoring forget requests), and initializes counters on indices for which CountSketch gives a frequency estimate of at least $\frac{1}{10}\eps'M^{(k)}.$  In order to achieve $\eps \norm{\ff}_2$ additive error we would like to remember checkpoints where the counter changes by an additive $\eps'\norm{\ff}_2$ amount. To do this, let $k_1, k_2, \ldots$ be the times at which $M^{(k)}$ achieves the values $1,2,2^2,\ldots.$  Between times $k_i$ and $k_{i+1}$ we store the time at which the counter was initialized, as well as the list of times at which the counter's value changes by $\eps' 2^i.$  At time $k_{i+1}$ we remove every other checkpoint, so that we maintain the counter's value to an additive $\eps' 2^{i+1}$ at all times.  Recall, that at any point in time we have at most $1/(\eps')^2$ active counters, so the total number of checkpoints stored is at most
\[
O\left(\frac{1}{(\eps')^2} + \sum_i \left\lfloor \frac{f_i}{\eps'\norm{\ff}_2} \right\rfloor\right).
\]
To bound the sum, note that at most $1/(\eps')^2$ terms are nonzero.  Let $\tilde{\ff}$ be the restriction of $\ff$ to the associated coordinates so that the sum is bounded by 
\[
\frac{\norm{\tilde{\ff}}_1}{\eps'\norm{\ff}_2} \leq \frac{\norm{\tilde{\ff}}_1}{\eps'\norm{\tilde{\ff}}_2} \leq \frac{1}{(\eps')^2}.
\]
So with no additional asymptotic space, we are able to store our checkpoints.  When we receive a prefix (or suffix) deletion request, we maintain an $\eps'\norm{\ff}_2$ approximation to the frequency count, simply by counting the number of checkpoints that occur after (or prior to) the deletion request.
\end{proof}

As a consequence, we obtain an algorithm for $F_p$ moment estimation in the prefix/suffix deletion models with the same guarantees as in the RFDS model. 

\begin{theorem}
\label{thm:F_p_PSD_model}
    There is an algorithm to obtain a $(1\pm \eps)$ approximation to $F_p$ for $p \geq 2$ in the $\alpha$-PSD model using space
$O\left(\frac{1}{(1-\alpha)^{2/p}}\frac{1}{\eps^{2 + 4 / p}} n^{1 - 2/p} \cdot \poly \log(n)\right)$.
\end{theorem}

\subsection{Entropy Estimation}
In this section, we consider estimating the entropy of the frequency vector $\ff$ in a data stream with forget operations, which is defined as $H(\ff) = -\sum_i p_i \log(p_i)$ where $p_i = |f_i| / \|\ff\|_1$. Here we assume $|H(\ff) - H(\widetilde{\ff})| \le \alpha H(\widetilde{\ff})$ and $F_1(\ff) \ge (1 - \alpha) F_1(\widetilde{\ff})$, where $\widetilde{\ff}$ is the vector without the forget operations and $\ff$ is the actual frequency vector with forget operations.. At a high level, \cite{HNO2008sketching} reduces additively estimating entropy to a multiplicative estimate of $F_p$ for $p$ close to $1$. Similarly to \cite{HNO2008sketching, KNW2011fast}, our algorithm will be based on our $F_p$ estimation algorithm. 

However, to utilize our $F_p$ estimation algorithm for some value of $p$, we are required to have the guarantee that $F_p(\ff) \geq (1-\alpha') \cdot F_p(\widetilde{\ff})$ for some $\alpha'$. To address this, we first show the following lemma. 
\begin{lemma}
    \label{lem:entropy}
    For $\alpha \in (0,1)$ and $p$ such that $|1-p| \leq O(\log n)$, assume that $|H(\ff) - H(\widetilde{\ff})| \leq \alpha H(\widetilde{\ff})$ and that $F_1(\ff) \geq (1-\alpha) F_1(\widetilde{\ff})$. Then we have
    \[
    F_p(\ff) \geq \left(1 - \alpha(p + (1 - p) H(\widetilde{\ff}))- O(\alpha^2 + (1 - p)^2 H(\widetilde{\ff})^2)\right)F_p(\widetilde{\ff}).
    \]
\end{lemma}
\begin{proof}
Observe that 
\[
\frac{F_p(\ff)}{F_p(\widetilde{\ff})} = \frac{\|\ff\|_1^p}{\|\widetilde{\ff}\|_1^p} \cdot \frac{\sum_i \left( \frac{f_i}{\|\ff\|_1} \right)^p}{\sum_i \left( \frac{{f'}_i}{\|\widetilde{\ff}\|_1} \right)^p}.
\]
Let us first consider the term \( \frac{\|\ff\|_1^p}{\|\widetilde{\ff}\|_1^p} \). Since we are given that \( \|\ff\|_1 \geq (1-\alpha)\|\widetilde{\ff}\|_1 \), it follows that
\[
\frac{\|\ff\|_1^p}{\|\widetilde{\ff}\|_1^p} \geq (1 - \alpha)^p \geq 1 - p\alpha - O(\alpha^2),
\]
where the last inequality is from a Taylor expansion.
Now consider the second term. Define \( p_i := f_i / \|\ff\|_1 \), and similarly \( p_i' := f_i' / \|\widetilde{\ff}\|_1 \). Let \( \Phi_p(\ff) := \sum_i p_i^p \), so that:
\[
\log \Phi_p(\ff) = (1 - p) H(\ff) + O((1-p)^2H(\ff)^2)
\]
and likewise for $\widetilde{\ff}$ via a Taylor expansion. 
Therefore,
\[
\frac{\Phi_p(\ff)}{\Phi_p(\widetilde{\ff})} = \exp( (1 - p)(H(\ff) - H(\widetilde{\ff})) + O((1 - p)^2 H(\widetilde{\ff})^2)).
\]
Given that \( H(\ff) \geq (1 - \alpha) H(\widetilde{\ff}) \), we have
\[
\quad
\frac{\Phi_p(\ff)}{\Phi_p(\widetilde{\ff})} \geq \exp( -\alpha(1 - p) H(\widetilde{\ff}) - O((1 - p)^2 H(\widetilde{\ff})^2) ) \geq 1-\alpha(1-p)H(\widetilde{\ff}) - O((1-p)^2H(\widetilde{\ff})^2).
\]
Combining both terms, we conclude:
\begin{align*}
\frac{F_p(\ff)}{F_p(\widetilde{\ff})} & \geq \left(1 - p\alpha - O(\alpha^2)\right) \cdot \left(1 - \alpha(1 - p) H(\widetilde{\ff}) - O((1 - p)^2 H(\widetilde{\ff})^2)\right) \\ 
&\geq 1 - \alpha(p + (1 - p) H(\widetilde{\ff}))- O(\alpha^2 + (1 - p)^2 H(\widetilde{\ff})^2). \qedhere
\end{align*}
\end{proof}

After obtaining \Cref{lem:entropy}, we can simply use our $F_p$ moment algorithms to get an entropy estimation. Specifically, the additive approximation entropy algorithm from \cite{HNO2008sketching} requires a $(1 + \eps')$-approximation to $F_p$ for $k = \log(1/\eps) + \log\log m$ different values of $p$ where $\eps' = \Theta\left(\frac{\eps}{k^3 \log m}\right)$. The values of these $p$ have the form $1 + y_i$ where $y_i$ is negative and $|y_i| \leq \frac{1}{2(k+1)\log m}$. Therefore, using our $F_p$ estimation algorithm from \Cref{thm:fpopt1}, we get an additive $\eps$ algorithm for estimating entropy using $\widetilde{O}\left(\frac{1}{\eps^2(1- t\alpha)} \cdot \poly \log m\right)$ where $t = 1 + O(1/\log m)$ and $m$ is the length of the stream. Recall that we have $H(\ff') \le \log n$. Then we have the following theorem.

\begin{theorem}
    \label{thm:entropy}
    Assume that $|H(\ff) - H(\widetilde{\ff})| \le \alpha H(\widetilde{\ff})$ and $F_1(\ff) \ge (1 - \alpha) F_1(\widetilde{\ff})$ for some $\alpha \in (0,1)$. There is a one-pass streaming algorithm that, with high constant probability, estimates the value of $H(\ff)$ within an $\eps$ additive error in the $\alpha$-RFDS model. This algorithm use $\widetilde{O}\left(\frac{1}{\eps^2(1- t\alpha)} \cdot \poly \log n\right)$ bits of space where $t = 1 + O(1/\log n)$.
\end{theorem}

\section{Lower Bounds for the Forget Model} \label{sec:LB}
The lower bounds below will use a direct sum result for information complexity several times. We recall a statement of this result here. This technique is standard in the literature and was introduced by Bar-Yossef, Jayram, Kumar, and Sivakumar~\cite{BJKS2004information}.  Theorem 2.1 of Molinaro, Woodruff, and Yaroslavtsev~\cite{MWY2013beating} proves a more general version of the fact below for two-player player games, although the same argument applies to any number of players.

\begin{proposition}
\label{prop:direct_sum_information_complexity}

Consider a $t$-player game with inputs $X_1, \ldots, X_t.$ with inputs $X = (X_1, \ldots, X_t) \sim \mu$. Further, suppose that there is a distribution $D$ such that the conditional distribution $X|D$ is a product distribution.

We say that the communication game has conditional information complexity at least $J$ on $X$, if for any correct protocol $\Pi,$
\[
I(\Pi ; X | D) \geq J.
\]

Now suppose that $(X,D_X)$ and $(Y,D_Y)$ are such that $X|D_X$ and $Y|D_Y$ product as above, with conditional information complexities $J_X$ and $J_Y.$  Let $\Pi_{X\times Y}$ be a protocol for the direct sum of $X$ and $Y$, which is a correct protocol for $X$ and $Y$ separately.  Then
\[
I(\Pi_{X\times Y}; X, Y | D_X, D_Y) \geq J_X + J_Y.
\]
\end{proposition}

\subsection{Lower Bound For $F_p$ for $p \in [0, 2]$}

We consider the following communication game which is a variant on Gap-Hamming.

\begin{problem}
\label{prob:gap_hamming_variant}
Alice holds $r$ one-hot binary vectors $\bv^{(1)}, \ldots, \bv^{(r)} \in \mathbb{R}^d$, each containing a single nonzero coordinate.  Bob holds binary vectors $\bw^{(1)}, \ldots, \bw^{(r)}$ in $\mathbb{R}^d$ as well as bits $b^{(1)}, \ldots, b^{(r)}.$  We are allowed one-way communication from Alice to Bob.  Let
\[
S = \sum_i b^{(i)}\langle \bv^{(i)}, \bw^{(i)} \rangle.
\]
With $3/4$ probability, Bob must distinguish between $S\geq \frac{r}{4} + C\sqrt{r}$ and $S\leq \frac{r}{4} - C\sqrt{r}.$
\end{problem}

We will reduce from the GapAndIndex problem introduced in Pagh, St\"{o}ckel, Woodruff~\cite{PSW2014min}. We restate the problem below. 

\begin{problem}
\label{prob:gap_and_index}
(GapAndIndex, \cite{PSW2014min}) Alice holds vectors $\bx^{(1)}, \ldots, \bx^{(r)}$ each in $\{0,1\}^d.$ Bob holds indices $i^{(1)}, \ldots, i^{(r)}$ in $[d],$ along with bits $b^{(1)}, \ldots, b^{(r)}.$  
Let
\[
S = \sum_{j=1}^r x^{(j)}_{i^{(j)}} \wedge b^{(j)}.
\]

After one-way communication from Alice to Bob, Bob must, with $3/4$ probability, distinguish between $S \geq \frac{r}{4} + C\sqrt{r}$ and $S \leq \frac{r}{4} - C\sqrt{r}$ where $C$ is an absolute constant.
\end{problem}

\begin{lemma}
\label{lem:gap_hamming_variant_lower_bound}
\Cref{prob:gap_hamming_variant} has an information complexity of $\Omega(r \log d)$ bits of information.  That is, if $\Pi = \Pi(\bv^{(1)}, \ldots, \bv^{(r)})$ is the transcript of a one-way protocol that solves GapAndIndex, then
\[
I(\Pi ; \bv^{(1)}, \ldots, \bv^{(r)}) 
\geq \Omega(r \log d).
\]
Moreover the hard distribution is a product distribution. That is, Alice and Bob can generate a hard distribution using private randomness and no communication.
\end{lemma}

\begin{proof}
We reduce from GapAndIndex.  Consider an arbitrary instance of GapAndIndex, where Alice holds $r$ vectors $\bx^{(1)}, \ldots, \bx^{(r)}$ each in $\{0,1\}^{\log d}$ (note that we have replaced $d$ with $\log d$ here), and Bob holds indices $i^{(1)}, \ldots, i^{(r)} \in [\log d],$ along with bits $b^{(1)}, \ldots, b^{(r)}.$
Alice interprets each of her vectors $\bx^{(j)}$ as a binary string representing a number $N^{(j)}$ in $[d].$  She then sets $\bv^{(j)}$ to be corresponding one-hot vector in $\{0,1\}^d$.

Let $S_j$ be the set of all numbers in $[d]$ whose $j$-th binary digit is $1.$  Bob sets $\bw^{(j)}$ to be the characteristic vector of the set $S_{i^{(j)}}.$
Note that $\langle \bv^{(j)}, \bw^{(j)}\rangle$ is precisely the $i^{(j)}$th binary digit of $N^{(j)}$, or in other words $x_{i^{(j)}}^{(j)}.$ It follows that
$\sum_i b^{(i)}\langle \bv^{(i)}, \bw^{(i)}\rangle$
is precisely the quantity $S$ from the GapAndIndex problem.

\cite{PSW2014min} states their bound as a communication lower bound, but in fact their proof gives an information lower bound of $\Omega(r\log d)$ for our choice of parameters.  Their hard distribution is for $\bx^{(j)}, i^{(j)}, b^{(j)}$ all uniform. In our setting, this translates to the hard distribution of $\bv^{(j)}$ uniform, $b^{(j)}$ uniform, and $\bw^{(j)}$ uniform over the sets $S^{(j)},$ which is clearly product.  (One could show that taking $\bw^{(j)}$ uniform over all binary vectors is also a hard distribution.)
\end{proof}

\begin{theorem}
\label{lem:lb_F0F1}
In the $\alpha$-RFDS model, computing a $1\pm\eps$ approximation to the $F_p$ moment for $p \in [0, 2]$ requires $\Omega(\frac{1}{1-\alpha}\frac{1}{\eps^2} \log n)$ space.
\end{theorem}

\begin{proof}
We first analyze the $O(1)$-RFDS model, and will later take a direct sum.  For this, we reduce from \Cref{prob:gap_hamming_variant} above. Suppose that we have an instance of this problem, with the same notation as in the problem statement.
We will set $d = \eps^2 n$ and $r = 1/\eps^2.$

Suppose that we have a sketch for $F_p$ moment estimation.  Alice computes the sketch $\bS_1$ on the vector $\bv = (v^{(1)}, \ldots, v^{(r)})\in \{0,1\}^n$ and sends it to Bob.  Let $\bw = (w^{(1)}, \ldots, w^{(r)}) \in \{0,1\}^n.$  Bob processes forget operations on all indices $j$ of $\bw$ with $w_j = 0.$ Bob also processes forget operations on all blocks $j$ with $b^{(j)} = 0.$

The resulting $F_p$ moments (note that the $F$ value will be equal for different $p$) are precisely the quantity $S$ from \Cref{prob:gap_hamming_variant}.  In our case a $1\pm \eps$ to the $F_p$ moment would yield an additive $\sqrt{\frac{1}{\eps^2}} = \sqrt{r}$ approximation.   Thus by \Cref{lem:gap_hamming_variant_lower_bound} we have an information lower bound of $\Omega(\frac{1}{\eps^2}\log(\eps^2 n)).$  Also note that the various forget operations only cause the $F_1$ moment to drop by a constant factor in expectation, so when $d$ and $r$ are sufficiently large constants, we are in fact in the $O(1)$-RFDS model with at least $9/10$ probability say, and thus will succeed at solving \Cref{prob:gap_hamming_variant} with $3/4$ probability.

\paragraph{Extending to the $\alpha$-RFDS model.} Since the lower bound of \Cref{lem:gap_hamming_variant_lower_bound} is information theoretic on a product distribution, the information complexity of the problem adds over independent instances. We take $\frac{1}{1-\alpha}$ copies of \Cref{prob:gap_hamming_variant} with the same parameters as above.  At the end of the stream, Bob attempts to solve all $\frac{1}{1-\alpha}$ problem instances.  To solve problem instance $k$, he performs forget requests on all problem instances other than instance $k.$  This yields $\frac{1}{1-\alpha}$ answers, one for each problem instance.  By the analysis in the paragraph above, the answer for each problem instance is correct with $3/4$ probability. Thus we have an $\Omega(\frac{1}{1-\alpha}\frac{1}{\eps^2}\log n)$ information lower bound, which implies the corresponding communication lower bound.
\end{proof}

\subsection{Lower Bound for $F_p$ Moments for $p\geq 2$}

We borrow from the proof by Woodruff and Zhou~\cite{WZ2021separations} which gives tight bounds for $F_p$ moment estimation in the non-forget model.  Their proof admits a direct sum as discussed above, which we use to extend to the $\alpha$-RFDS model.

\begin{theorem}
\label{thm:lb_Fp}
A streaming algorithm for solving solving $F_p$ moment estimation to $(1\pm \eps)$ multiplicative error in the $\alpha$-RFDS model requires $\Omega(\frac{1}{(1-\alpha)^{2/p}} \frac{1}{\eps^2} n^{1 - 2/p})$ space.
\end{theorem}

\begin{proof}
\cite{WZ2021separations} introduces a problem that they call $(t,\eps, n)$-DisjInfty. We give their formal definition. 
\begin{definition}
   In the $(t,\eps, n)$-player set disjointness estimation problem $(t, \eps, n)$-DisjInfty, there are $t+1$ players $P_1, \ldots, P_{t+1}$ with private coins in the standard blackboard model.  For $s \in [t]$, each player $P_s$ receives a vector $\bv_s \in \{0,1\}^n$ and player $P_{t+1}$ receives both an index $j \in [n]$ and a bit $c \in \{0,1\}$. For $u = \sum_{s \in [t]} \bv_s$, the inputs are promised to satisfy $u_i \leq 1$ for each $i \neq j$ and either $u_j = 1$ or $u_j = t$. With probability at least $9/10$, $P_{t+1}$ must differentiate between three possible input cases: 
   \begin{enumerate}
       \item $u_j + \frac{ct}{\eps} \leq t$
       \item $u_j + \frac{ct}{\eps} \in \{\frac{t}{\eps}, \frac{t}{\eps} + 1\}$
       \item $u_j + \frac{ct}{\eps} = (1+\eps)\frac{t}{\eps}$,
   \end{enumerate}
   where $\eps \in (0,1)$. 
\end{definition}

They then prove the following. 
\begin{lemma}(Theorem 4.18, \cite{WZ2021separations})
    The total communication complexity for the $(t, \eps, n)$-player set disjointness estimation problem is $\Omega(n/t)$. 
\end{lemma}

Moreover, their proof of Theorem 4.18 gives a conditional information lower bound with respect to a distribution $D$, conditioned on which the players' inputs are drawn from a product distribution.  This, along with their lower bound, implies that the direct sum of $N$ copies of the $(t,\eps,n)$-DisjInfty has a communication lower bound of $\Omega(N n/t).$

The reduction given in the Theorem 4.19 of \cite{WZ2021separations} shows that solving $F_p$ moment estimation for a vector of length $n$ and $F_p$ moment $\Theta(n)$ requires $\Omega(\frac{1}{\eps^2} n^{1 - 2/p})$ space via a direct reduction from the $(t,\eps, n)$-DisjInfty problem.  

Now suppose we have an algorithm for $F_p$ moment estimation in the $\alpha$-RFDS model for streams of length $nN$, where $N = \frac{1}{1-\alpha}$.  Player $t+1$ can solve a single instance of the $(t,\eps, n)$-DisjInfty with constant advantage, simply by performing the reduction to $F_p$ estimation on each instance, and then issuing forget requests on all but $1$ instance.  By performing the procedure locally $N$ times (which requires no communication), all $N$ instances are solved with constant advantage.

Thus for universes of size $nN$ in the $\alpha$-RFDS model, we have a lower bound of $\Omega(\frac{1}{1-\alpha}\frac{1}{\eps^2} n^{1 - 2/p})$ for $F_p$ moment estimation.  Replacing $n$ with $n/N = (1-\alpha)n$ gives our stated lower bound.
\end{proof}

\subsection{Turnstile Streams}
Finally we give a simple lower bound against turnstile streams (and even strict turnstile streams) in the $\alpha$-RFDS model which justifies our focus on insertion-only streams.
Recall that a turnstile stream allows insertions and deletions. Specifically, updates are of the form $(i, +1)$ or $(i, -1)$ which performs $f_i = f_i + 1$ and $f_i = f_i -1$ respectively. In the strict turnstile model we are guaranteed that deletion is never applied to a coordinate of $\ff$  that is currently zero. 

We now introduce the following problem to prove the lower bound. 

\begin{problem}
\label{prob:three_way_game}
There are three players, Alice, Bob, and Charlie.  Communication is one-way from Alice $\rightarrow$ Bob $\rightarrow$ Charlie.  Alice and Bob each hold subsets $A$ and $B$ of $[n]$ that are promised to intersect in a single element.  Charlie is given a copy of $A$ as well as two different values $x_1,x_2 \in [n]$, one of which is promised to be the common element of $A$ and $B.$  Charlie must decide with $3/4$ probability which of the two elements is common to $A$ and $B.$
\end{problem}

\begin{lemma}
\label{lem:three_way_game_lower_bound}
\Cref{prob:three_way_game} requires $\Omega(n/\log n)$ communication.
\end{lemma}

\begin{proof}
We reduce from the set-disjointness problem, with the promise that the intersection size is either $0$ or $1.$ This problem is known to have an $\Omega(n)$ lower bound for two player protocols (see \cite{BJKS2004information} for example).

In this problem Alice and Bob holds subsets $A, B\subseteq [n]$ respectively and must decide if $|A\cap B|$ is $0$ or $1.$ We show that our three-player game solves this problem.  First observe that by running our $3$-player game on $O(\log n)$ independent instances and taking a majority vote, we may obtain a protocol for the three-player game with failure probability at most $\frac{1}{4n}.$

Now suppose that we have such a protocol. To solve set-disjointness we run the three-player protocol, however Bob sends his sketch back to Alice, rather than to Charlie.  Now Alice runs Charlies protocol on all pairs of distinct elements $(x,y)\in [n]\times[n].$ 

Suppose that $A$ and $B$ are not disjoint with $A\cap B = \{z\}$. Since the failure probability for the three-player game was reduced to $\frac{1}{4}\frac{1}{n^2},$ with $3/4$ probability Alice (while simulating Charlie's protocol) produces the output $z$ on all pairs containing $z.$  There can can only be one such element with the property, so Alice sends it to Bob to check whether it lies in $B.$  Thus with $3/4$ probability when $A$ and $B$ are not disjoint, Alice and Bob produce a proof of this fact (namely the element $z$ which they both verify are in their respective sets).  Otherwise, if the protocol fails to produce such a proof, then they output that the sets are disjoint.

Note that this protocol guesses randomly half the time, but otherwise succeeds with $9/10$ probability, and therefore succeeds with at least $7/10$ probability.  Repeating a constant number of times boosts this to a $3/4$ probability of success.
\end{proof}

\begin{theorem}
\label{thm:lb_turnstile}
In the $\frac{2}{3}$-forget model with strict turnstile updates, approximating either the $F_0$ moment, or any $F_p$ moment with $p\geq 1$ to relative error $1\pm\frac{1}{10}$ with $9/10$ probability requires $\Omega(n/\log n)$ space.
\end{theorem}

\begin{proof}
We give two protocols that give a valid reduction for different ranges of $p$. The second will be a small variant on the first.

\paragraph{$p=0$ and large $p$ protocol.}  Suppose for now that $p\geq 1.25.$ We observe that a turnstile streaming algorithm in the $O(1)$-forget model could solve \Cref{prob:three_way_game}. Alice simulates the streaming algorithm, and performs insertions for all elements of $A,$ then sends the result sketch to Bob.  Bob performs a forget operation for all elements of $B$, and sends the sketch to Charlie.  Finally, Charlie performs a turnstile removal update on all elements of $A\setminus\{x_1, x_2\}.$  Note that Charlie only removes elements in $A\setminus B$ by the promise, so this is valid for a strict turnstile algorithm.  Finally, Charlie executes an insert operation on $x_1.$

If $A\cap B = \{x_1\}$, then at the end of the stream Charlie holds a sketch of a vector $\bx$ whose support has support entries are precisely the multiset $\{2\}$.  On the other hand if $A\cap B = \{x_2\}$ then $\bx$ has support entries $\{1,1\}.$  For any fixed $p \geq 1.25$, an algorithm that produces a $1 \pm \frac{1}{10}$ multiplicative approximation to $F_p$ would distinguish these two instances.  The lower bound now follows from \Cref{lem:three_way_game_lower_bound}.

\paragraph{Small $p$ protocol.} Suppose that $p \leq 1.25$. we prove the lower bound for on a universe of size $2n.$  Denote the elements of this inflated universe as $1,2,\ldots, n, 1', \ldots, n'.$  Using shared randomness, the players first choose a subset $S$ of $[n]$ where each element of $[n]$ is included pairwise independently with probability $1/2.$  Note that this choice of $S$ may be implemented with a pairwise independent hash function, and so may be communicated between the players with only an additive $O(\log n)$ number of bits.

We run the protocol from above, except that for each item $i$ in $S$ the corresponding update is performed on both $i$ and $i'.$  Also if $x_1$ and $x_2$ are either both in $S$, or both not in $S$ then, Charlie will simply choose to output $x_1$ with $1/2$ probability and $x_2$ with $1/2$ probability.  

Suppose that this does not occur and that $x_1 \in S$ and $x_2 \notin S.$
If $A\cap B = \{x_1\}$ then the entries of the final support are $\{2,2\}.$  If $A\cap B = \{x_2\}$ then the entries of the support are $\{1,1,1\}.$  Thus Charlie can distinguish between these two cases via a $(1 \pm \frac{1}{10})$ to $F_p.$ 

In the other case where $x_1 \notin S$ and $x_2 \in S$, the final support is either $\{2,0\}$ or $\{1,1,1\}$ and again Charlie can distinguish between the two with a $(1\pm \frac{1}{10})$ approximation to $F_p.$

Finally, note that the overall success probability is at least $\frac{1}{2}(\frac{1}{2} + \frac{9}{10}) \geq 0.7,$ and the success probability can be boosted to $3/4$ with a constant number of repetitions.

\end{proof}

\bibliographystyle{alpha}
\bibliography{general}

\end{document}